\documentclass[referee,usenatbib]{iupartex}
\usepackage{newtxtext,newtxmath}
\usepackage[T1]{fontenc}
\usepackage{ae,aecompl}

\usepackage{multicol}   
\usepackage{bm}		    
\usepackage{pdflscape}	

\title[Age-metallicity relation]{The Age-Metallicity Relation in the Solar Neighbourhood}

\author[D\"oner et. al]{%
S. D\"oner $^{1\cc}$\orcid{0000-0001-7922-8961},
S.\,Ak$^{2}$\orcid{0000-0002-0912-6019} 
\"O.\,\"Onal Ta\c s$^{2}$\orcid{0000-0003-0864-1921}
and
O. Plevne$^{2}$ \orcid{0000-0002-0435-4493}
\affsep \\
$^1$Istanbul University, Institute of Graduate Studies in Science, Programme of Astronomy and Space Sciences, 34116, Beyaz{\i}t, Istanbul, Turkey\\
$^2$Istanbul University, Faculty of Science, Department of Astronomy and Space Sciences, 34119, Beyaz\i t, Istanbul, Turkey\\
}

\corres{S. D\"oner}{sibeldnr@gmail.com}

\pubyear{2023}

\doiheader{XXXXXXX/PAR.20XX.00000}
\date{
	\pSubmit{00.00.0000} 
	\pRevReq{00.00.0000}
	\pLastRevRec{00.00.0000}
	\pAccept{00.00.0000}
	\pPubOnl{00.00.0000}
}
\volume{0}
\volnumber{1}

\nolinenumbers 

\begin{document}
\label{firstpage}
\pagerange{\pageref*{firstpage}--\pageref*{lastpage}}
\maketitle

\begin{abstract}
Age-metallicity relation for the Galactic disc is a crucial tool and to constrain the Galactic chemical evolution models. We investigate the age-metallicity relation of the Galactic disc using the red giant branch stars in the Solar neighbourhood. The data cover the Galactocentric radius of $7\leq R_{\rm gc} (\rm kpc) \leq9.5$, but extends up to 4 kpc in height from the Galactic plane. We use kinematic age derived from highly precise astrometric data of {\it  Gaia} Data Release 2 and element abundance ratios from high-resolution spectroscopic data of APOGEE-2 catalogues. We apply a two-component Gaussian mixture model to chemically separate the programme stars into thin and thick disc populations. The stars in each population are grouped into different distance intervals from the Galactic plane. The mean metal abundances and velocity dispersions of the stars in the groups were calculated and the kinematic ages were determined from their kinematic parameters. We found a steep relation for the thin disc with -0.057$\pm$0.007 dex Gyr$^{-1}$, and even a steeper value of -0.103$\pm$0.009 dex Gyr$^{-1}$ for the thick disc. These age-metallicity relations along with the prominent differences in age, metallicity, and kinematic behaviours seen from the data, clearly show it is important to consider the distinct formation scenarios of the Galactic disc components in modelling the Milky Way.
\end{abstract}




\section{Introduction}
\label{sec:Introduction}
Galactic archaeology is built upon to dependable features of stars, and interstellar medium to reconstruct the formation history by considering certain evolutionary scenarios. Features like  kinematics, age and chemistry help to track the past events of the Galactic formation. It is known that kinematics of a star can change over time due to large scale perturbations and adapt the surrounding environment. Age of a star can be estimated from the atmosphere model properties using statistical methods or it can be inferred from the kinematic properties once the basic observational parameters are determined precisely. Chemistry of a star changes as a  function of time during the Galactic evolution. From the first concrete formation scenarios of ELS \citep{ELS62} and SZ \citep{SZ78} to sophisticated chemo-dynamic models \citep[i.e.][]{Minchev13}, our understanding of the Milky Way has evolved with the aid of precise photometric, spectroscopic and astrometric data of hundreds of millions of stars where most of them lie within the extended Solar neighbourhood ($<2$ kpc). This knowledge has enabled us to get a rough idea of the chemical and kinematic behaviours of the stars as we move away from the Galactic centre in all directions within the Milky Way. Accordingly, the stars in the central bulge region are old and their metallicity tends to increase radially away from the Galactic centre and decrease perpendicularly from the Galactic plane. This suggests that the outer and central parts of the Milky Way were formed earlier than the disk objects \citep{ELS62}. This general age trend is followed by inversely related iron and alpha-element abundance ratios, such that older stars have lower iron and higher alpha abundances, as a result of Type II supernova activity, than younger ones. On the other hand, atmospheric abundances of younger stars indicate that iron and alpha element abundance ratios are  reversed, as a consequence of increasing  Type Ia supernova activity in our Galaxy. This shows that there is a way to track the evolution of the Milky Way using chemical element abundances, which is the basis of the chemical evolution models. 

Chemical evolution has been one of the ways to understand the evolution of the Milky Way disc since its formation. Trapped in stars the primordial gas is changed with stellar evolution. Since the Galaxy evolution is taking a lot longer time than stellar evolution, the gas in the Galaxy become a subject to a recycled chemical enrichment over time. The formation mass of a star determines the chemical element production yield and the method of recycling of the gas. At one end, if the star is a low-mass star than the gas is trapped inside the star will hold it's initial chemical variety for a long time. At the other end if the star is massive, then star will produce a variety of chemical elements via nuclear reactions and return the gas with significant change in chemistry. The general effect of this recycling process can be resolved with the accurate determination of the star formation history. This mechanism is responsible for increasing metallicity in stars, thus also a measure of time evolution. Elements leave their imprints, which are measurable, on the star light through spectral measurements. This interpretation has already been made by several authors as chemical tagging \citep{Freeman02} or the chemical clock approach \citep*{CMG97}. Since the distribution of metallicity has a tendency to evolve in time due to differences of initial mass function across the Galaxy and due to spatial velocities of stars. This provides a unique way to infer the star formation rate, metallicity distribution, amount of gas accretion and supernova activity. Within the last decade, chemical evolution models have been combined with dynamic models to form chemo-dynamic ones, giving greater insight into the Galaxy formation \citep{Minchev16}.

The last two decades have witnessed major developments in observational astronomy that enlarge our understanding of the Milky Way. The first decade of 2000's were dominated by photometric all sky surveys that prepared the necessary grounds to initiate spectroscopic sky surveys such as GCS \citep{Nordstrom04}, RAVE \citep{Steinmetz06}, BRAVA \citep{Rich07}, APOGEE \citep{Allende08}, SEGUE \citep{Yanny09}, GES \citep{Gilmore12}, LAMOST \citep{Zhao12}, and GALAH \citep{deSilva15}. These surveys made it possible to study the structure, formation and evolution of the Milky Way in depth. Galactic disc formation and evolution studies rely on the observations and theoretical models by means of chemistry, kinematic properties and ages of its main ingredients, i.e. stars, gas and dust. Among these properties the chemical content of up to 20+ elements in the atmospheres of stars in the Solar neighbourhood is obtained more reliably with the aid of high resolution and high signal-to-noise spectroscopic all sky surveys. Kinematic properties of stars can be derived by five astrometric parameters and one spectroscopic parameter, i.e. radial velocity, in which the astrometric ones are now provided by the {\it Gaia} \citep{Gaia18}.

The first age-metallicity relation (AMR) studies used the {\it Hipparcos} data of F-G spectral type main-sequence stars. Using these data, \cite{Twarog80} speculated for the first time that there might be a relation between age and metallicity. \cite{Edvardsson93} suggested that there exists a wide spread between age and metallicity that implies there may not be such a relation between these parameters. \citet*{Carraro98} analysed the AMR for four different data types, i.e. F-G main-sequence stars, open clusters, synthetic populations and chemical evolution models, then compared the resulting AMRs. In the study, ages were calculated with the synthetic colour-magnitude diagram method using the metallicity, colour excess and distance modulus, while metallicities are determined spectroscopic and/or photometric methods, which the data regarding is compiled from independent sources in the literature. They pointed out that found AMRs are unable to explain the presence of old and relatively metal-rich stars and clusters. They found there is a large spread in metallicity at any age. In another study, \cite{Ng98} determined a gradient of the AMR as -0.07 dex Gyr$^{-1}$. In studies that use combined astrometry with Str\"omgren photometry data of Solar neighbourhood stars \citep{Olsen83, Olsen93}, \citet*{Feltzing01} found a prominent AMR for a sample of early spectral type stars while no AMR for late-type stars by asserting these stars which have longer lifetime. \citet{Nordstrom04} have used the data consisting of around 14,000 main-sequence stars, but despite the sensitivity of their data they claimed that their calculated AMR was dispersed because of the uncertainties. \citet{Soubiran08} showed no relation between age and metallicity for red giant branch (RGB) stars older than 5 Gyr. Using RAVE DR3 spectroscopic data of F-G spectral type main-sequence stars, \citet{Duran13}investigated the age and metallicity relation. They have shown that the AMR is related to the spectral type, but not to the population type. \citet{Haywood13} have studied the AMR in the Galactic disc and found -0.025 dex Gyr$^{-1}$ for the thin disc and -0.15 dex Gyr$^{-1}$ for thick disc sample. \citet{Bergemann14} used the high-resolution UVES spectra of GES survey to investigate relationship between age, metallicity and $\alpha$-enhancements in F-G main-sequence stars. They claimed a nearly flat AMR for stars younger than 8 Gyr, while no AMR for stars older than 8 Gyr due to spread in metallicity in the Solar neighbourhood.

\citet{Lin18} have investigated the AMR using 4,000 stars selected from {\it Gaia} DR1 TGAS catalogue. They have found -0.032 dex Gyr$^{-1}$ of AMR gradient for stars older than 10 Gyr. \citet{Wojno18} have analyzed 25,000 FGK spectral type main-sequence stars in $7<R_{\rm gc}({\rm kpc})<9$ selected from RAVE and TGAS catalogues by separating their sample into low- and high-$\alpha$ sub-samples. This study claims the existence of AMR in both chemical discs. \citet{Feuillet19} also supports this result. In their study, they have investigated the AMR for iron and $\alpha$-element abundance ratios using the RGB stars selected from APOGEE catalogue and {\it Gaia} DR2 astrometry. They have claimed that there is a significant variation between age and metallicity as a function of current Galactocentric radial and vertical distances from the Galactic centre and Galactic plane, respectively, for the first time. They have suggested that AMR supports the radial migration of stars on the disc plane. 

In recent studies, AMR investigations have become more intricate than before due to the contribution of additional element abundance ratios, i.e. iron-peak, alpha-, neutron-capture and s-process elements, from HRS spectra. \citet{Casali20} have studied relations between [X/Fe] and stellar age for 560 solar-like stars in the Solar neighbourhood using their HARPS spectra. They have shown that there are strong correlations between [X/Fe] and stellar age with negative slope for s-process elements, a positive slope for $\alpha$-elements and nearly flat relation for iron-peak elements. \citet{Nissen20} have analysed the relations between 13 element abundance ratios and model ages of 72 solar-like stars with HARPS spectra. They found two sequences in the age-metallicity distribution and have interpreted these as an evidence of gas accretion in two episodes onto the disc. However, they have concluded that there is a need for a deeper study that eliminates the systematic errors in abundance derivation.

 An important issue in AMR studies is the estimation of accurate stellar age, which is not a directly measured property. There are various methods to obtain this information. The most popular one depends on the isochrone matching based on the application of Bayes statistics to the observational priors defined from the stellar spectra. This application gives a probability density function \citep{Pont04, Jorgensen05} which helps to define the most likely age. This method provides relatively precise ages \citep[$\pm$1 Gyr;][]{Edvardsson93, Casagrande11, Bensby14} for stars in middle spectral types (FGK), because it depends on the precision of stellar evolution tracks and only iron abundance ratios, so far.  Theoretical tracks, even though they are improved over time, give better results for relatively metal-rich stars. At the upper, and lower limits of the metallicity range $-2.2\leq {\rm [Fe/H] (dex)}\leq 0.50$ for PARSEC; $-4\leq {\rm [Fe/H] (dex)}\leq 0.45$ for MIST for the Galactic studies, probability density functions are impaired so that stellar age cannot be determined, especially for stars with [Fe/H]$\leq$ -2.2 dex. Another age determination method uses astroseismology. However this method is not applicable to all stars. As a star evolves it's chemical composition changes and this affects the average mass of a particle in a gas, so that the pressure at a given temperature. This results in a change in sound speed. This method measures the sound speed inside a star by finding the characteristic oscillation frequencies, which gives the age of the star.

Old AMR studies have used early and late spectral type main-sequence stars. The stars spend 90\% of their lifetimes in the main sequence stage. Especially, atmospheres of FGK main-sequence stars reflect the chemical properties of their proto-stellar cloud, so the most reliable metallicity can be inferred from these objects. These objects are fairly common stars in the Solar neighbourhood, since they were targeted in most of the spectroscopic sky surveys. However, these stars cannot provide information from great distances due to their relatively low luminosity. Nevertheless, AMR can be probed in great distances using red giant branch (RGB) stars as stellar tracers instead of main-sequence stars. RGBs are bright objects that reside on the red, metal-rich part of the observational colour-magnitude diagram. An RGB star has an isothermal non-burning helium core with degenerate electrons \citep{Chiosi98} and produces core energy from the surrounding shell of it's core and there is a large convective envelope beyond this shell. Red giants are evolved stars which "burn" hydrogen in a shell around an inert helium core (see Iben 1968). If the  mass of the star is sufficiently large, then this core grows massive enough to initiate helium fusion \citep{Schwarzschild_1962}. The helium mass in the core steadily increases as the shell feeds the core with freshly synthesized helium and this cause the star to steadily shine brighter in time and steadily decrease in effective temperature. RGB stars lose mass from their convective envelopes and this cause dust shells to form around them \citep{Origlia02, Boyer08}. This means that age estimation using stellar isochrones with spectroscopic priors becomes an impossible task for RGB stars. On the other hand, HRS spectroscopic data of APOGEE-2 and highly precise {\it Gaia} astrometric data helps to find other solutions for RGB stars. Spectroscopic and astrometric data can be used to determine the space velocity components and especially their dispersions, which will help to determine stellar ages kinematically \citep{Wielen77}.

In this study, we present the results on the AMR derived from the RGB stars in the Solar neighbourhood that belong to the different chemical Galactic disc components. Section 2 explains the data selection, distance and space velocity estimation. Section 3 describes the chemical separation of RGB stars into disc components and kinematic age derivation. Section 4 deliberates the results for calculated AMRs of chemical discs and Section 5 gives the summary and discussion of the results by comparing with the literature.

\section{Data}
\begin{figure*}
\centering
\includegraphics[width=\columnwidth]{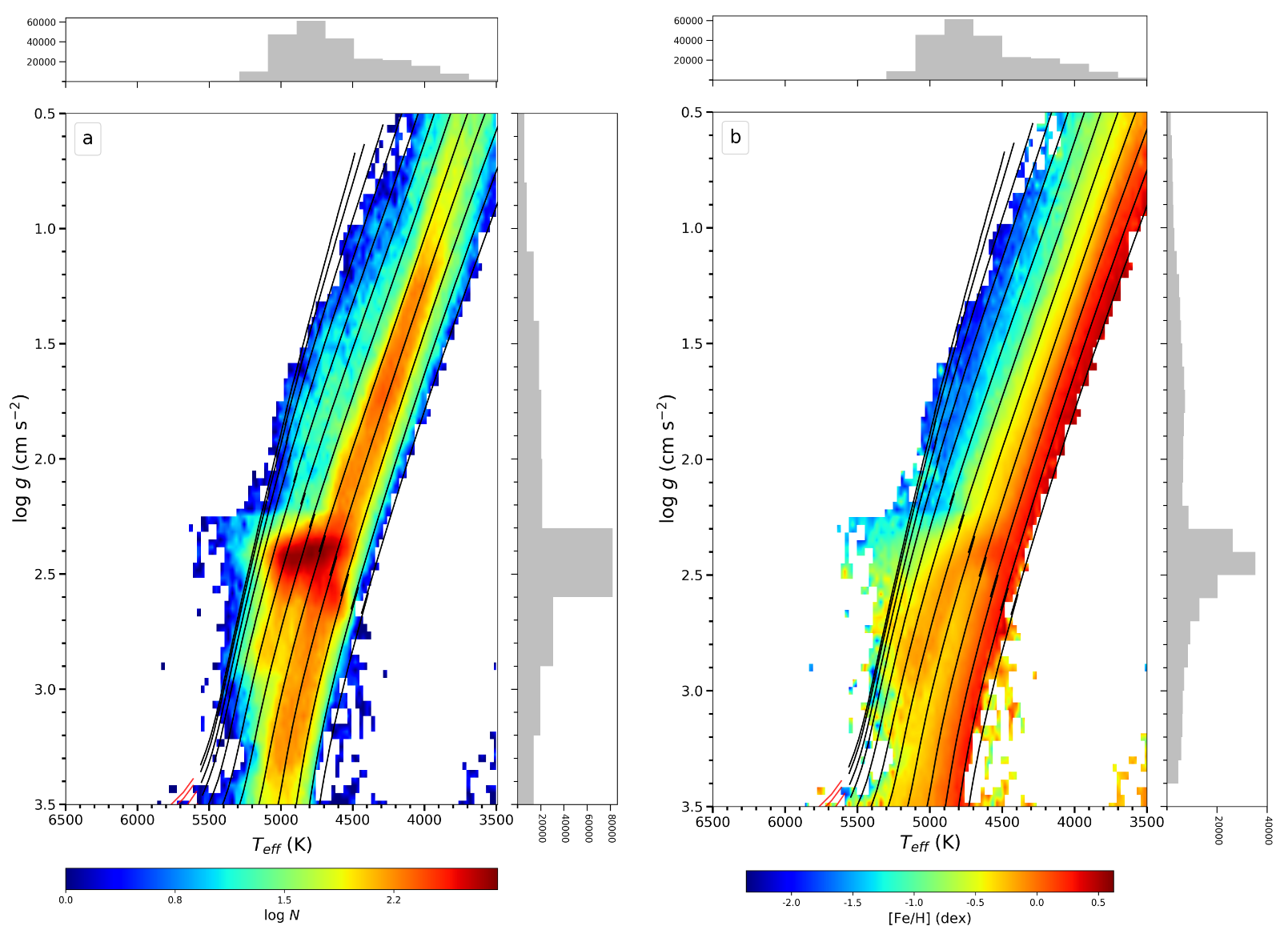}
\caption{Kiel diagrams of the RGB stars colour coded for the stellar number density (a) and the metallicity (b), respectively. $T_{\rm eff}$ and $\log g$ histograms are plotted parallel to apses and ordinate axes in both HR diagrams.} 
\label{fig:Fig1}
\end{figure*}

The AMR for the Milky Way disc is investigated using the Sloan Digital Sky Survey’s (SDSS) sixteenth data release (DR16) \citep{Blanton17, Ahumada20} taken as part of the second, dual hemisphere phase of APOGEE (APOGEE-2) \citep{Majewski17}. The APOGEE/DR16 dataset includes about 473,307 stars using the two 300-fibre APOGEE spectrographs \citep{Wilson19} and the APOGEE survey has near-complete coverage in Galactic longitude, due to the first release of data from Las Campanas Observatory in Chile \citep{Bowen73}. The APOGEE spectroscopic data reduction pipeline \citep[cf][]{Holtzman15, Nidever15} supply radial velocity ($\gamma$) and atmospheric model parameters ($T_{\rm eff}$, $\log g$, [Fe/H]), while more than 20 different elemental abundances are provided from the APOGEE Stellar Parameters and Chemical Abundances Pipeline \citep[ASPCAP;][]{Holtzman18, GarciaPerez16}. APOGEE/DR16 data is extracted from high resolution ($R\sim 22,500$) and high  signal-to-noise ($S/N>100$) spectra and complemented with the second data release of {\it Gaia} \citep[Gaia DR2;][]{Gaia18} astrometric and photometric data. So, no additional cross-match is performed between APOGEE-2 and {\it Gaia} DR2 catalogues. There are 407,266 stars out of 473,307 have all atmosphere model parameters ($T_{\rm eff}$, $\log g$, [Fe/H], [$\alpha$/Fe]) and trigonometric parallaxes ($\varpi$) in the APOGEE-2 catalogue. These stars cover $3000<T_{\rm eff}{\rm (K)}<9000$ and $1<\log g~{\rm (cgs)}<6$ intervals on the $\log g \times T_{\rm eff}$ plane. Stars with negative parallax values are also eliminated from the sample. In this study, AMR is investigated using RGB stars, which are selected by constraining the data on the effective temperature and the surface gravity plane, i.e. $3500<T_{\rm eff}{\rm(K)}<6500$ and $0.5<\log g~{\rm (cgs)}<3.5$. RGB stars are colour coded by the logarithmic number density (left panel) and the metallicity (right panel) as shown in the Fig. \ref{fig:Fig1}. In this sample, there are 233,568 objects with $\rm {[Fe/H]}\geq -2.5$ dex and $\varpi>0.1$ mas. PARSEC isochrones \citep{Bressan12} are also plotted on the $\log g \times T_{\rm eff}$ plane by assuming constant stellar age of 7 Gyr and the PARSEC isochrones are chosen for $-2.25<{\rm [Fe/H] (dex)} <0.5$ metallicity range with 0.25 dex steps. These isochrones are plotted on the Kiel diagram with black-solid lines and it can be seen that these calibrated theoretical models are in good-agreement with observational data. Also in the figure, the $T_{\rm eff}$ and $\log g$ histograms are presented parallel to apses and ordinate axes in both Kiel diagrams. RGB stars in the APOGEE-2 survey have high $S/N$, so the $S/N$ median values at 50\%, 68\% and 95\% of the RGB samples are 175, 125, 75, respectively. The precision in APOGEE-2 radial velocities of RGB stars is around $\leq 2$ km s$^{-1}$, so there is no need to put constraints on both $S/N$ and radial velocity.

Our RGB sample intersects with relatively metal rich red clump stars, which are easily spotted on the Fig. \ref{fig:Fig1}'s panel (a) as a dark red dense elipsoidal region. This region is a mixture of ascending red giants, descending red giants after helium flash and stable helium burning red clump stars. We used the technique of of \cite{Plevne20} to determine the contamination of this region by red giant at different level of stellar evolution. In this technique RC stars were determined using the width of half maximum and the standard deviation of the Gaussian distributions on the $\log g \times T_{\rm eff}$ plane plane. Based on this analysis it is found that RGB sample is contaminated only 22$\%$ with RC stars according to the 1$\sigma$ of the distribution of stars within the ellipsoid.

\begin{figure}
\centering
\includegraphics[width=0.6\columnwidth]{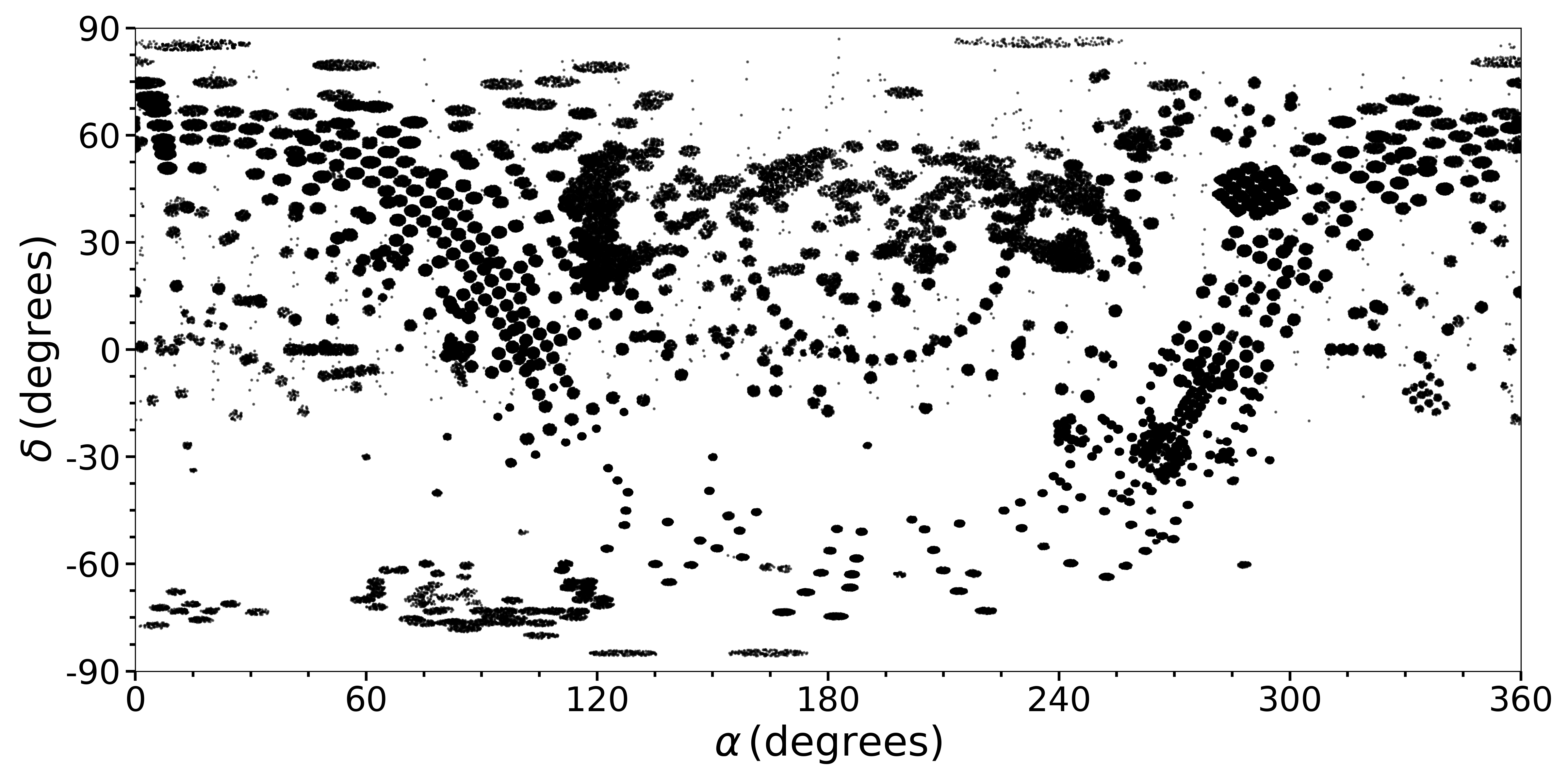}
\includegraphics[width=0.6\columnwidth]{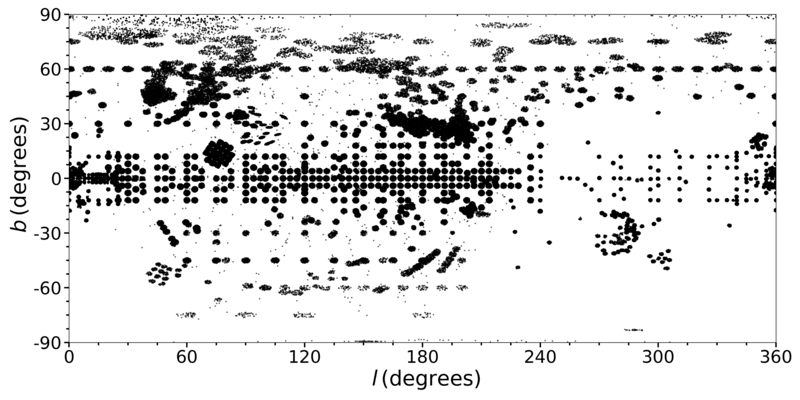}
\caption{The positions of 233,568 RGB stars in Equatorial (upper panel) and Galactic (lower panel) coordinate systems.}
\label{fig:Fig2}
\end {figure} 

\begin{figure}
\centering
\includegraphics[width=\columnwidth]{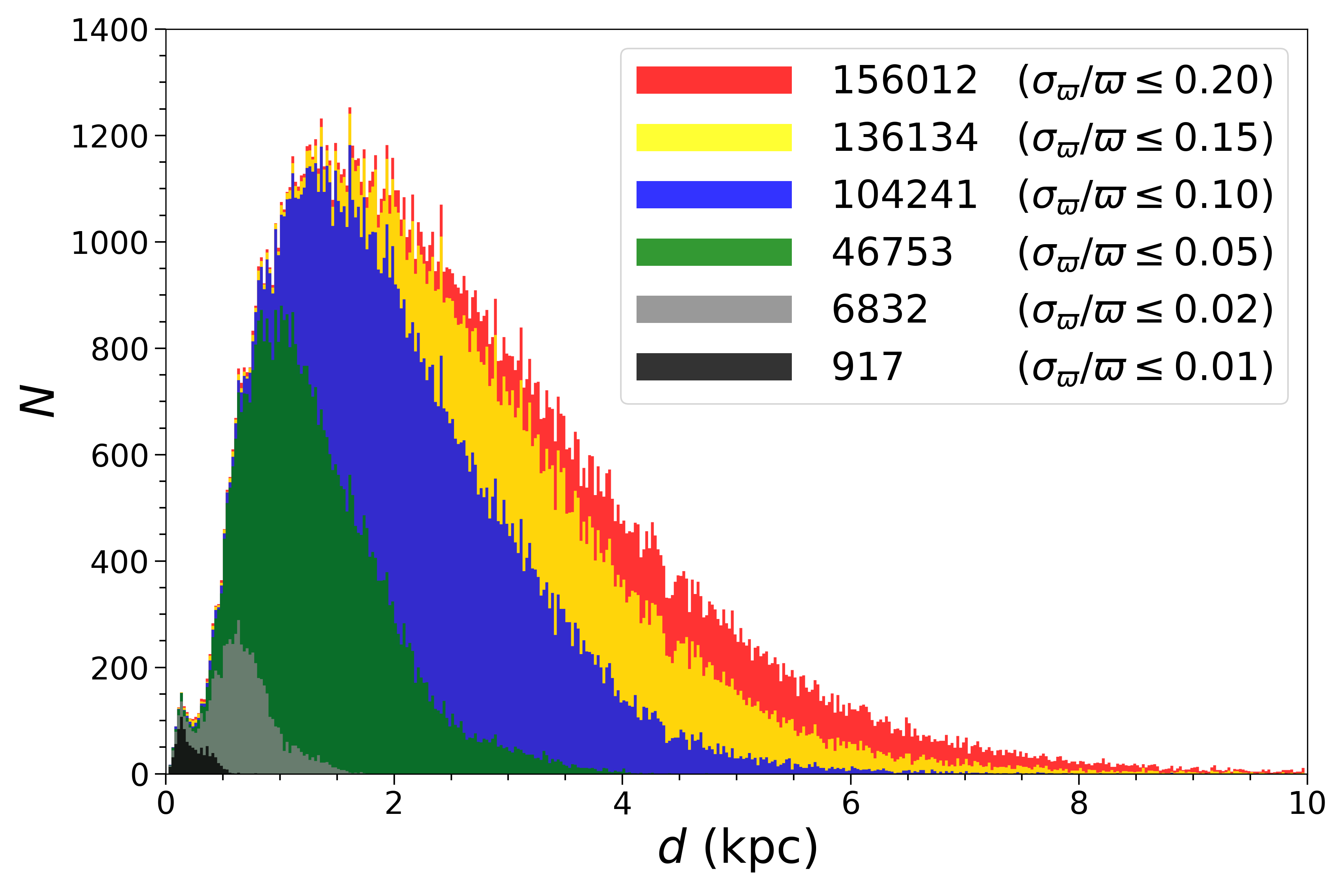}
\caption{Distance histograms of RGB stars based on different relative parallax errors.}
\label{fig:Fig3}
\end {figure}

The position of 233,568 RGB stars in equatorial and Galactic coordinates are shown in Fig. \ref{fig:Fig2}. Most of the sources reside on the northern hemisphere. In the southern hemisphere there are relatively small number of sources observed within the APOGEE-2 campaign. Many of the APOGEE-2 stars are observed on the Galactic plane which is the main difference from the other spectroscopic sky surveys. Unlike the position of stars in equatorial coordinates, there are more stars in northern Galactic hemisphere in Galactic coordinates, which is caused by the 63$^{\circ}$ difference between two coordinate systems. It seems the regions where $|b|>20^{\circ}$ are surveyed more systematically. There are 142,178 stars in the northern and 91,390 stars in the southern hemisphere. Distance histograms of APOGEE-2 RGB stars are shown for various relative parallax error limits in Fig. \ref{fig:Fig3}. The sub-samples are obtained for $\sigma_{\varpi}/\varpi\leq 0.20$, 0.15, 0.10, 0.05, 0.02, 0.01 and the number of stars is given in the figure. Percentage of the RGB stars corresponding to above relative parallax error limits are 67\%, 58\%, 45\%, 20\%, 3\% and 0.4\%, respectively. So it is preferred to limit the relative parallax errors to 0.1 in order to minimize the errors in distance estimation, which reduces the sample size to 104,241 stars.

The RGB sample is further divided into the discrete relative parallax error intervals at 0.05, 0.08 and 0.10, but this time to investigate the dispersed effect and the level of necessity of applying the \citet[][LK]{Lutz73} bias correction to the {\it Gaia} DR2 parallaxes. These intervals include 45\%, 37\% and 18\% of the APOGEE-2 RGB samples, respectively. For these discrete sub-samples, the LK corrections, to be applied to the {\it Gaia} DR2 trigonometric parallaxes that are calculated using  Eq. 12 of \citet{Smith87}, are found as $\leq 1\%$, 1-2.6\% and 2.6-4.3\% respectively \citep[see also,][]{Celebi19}. Based on these results no LK correction is applied to the trigonometric parallaxes of 104,241 RGB stars.

\subsection{Stellar Distances}
\begin{figure}
\centering
\includegraphics[width=0.60\columnwidth]{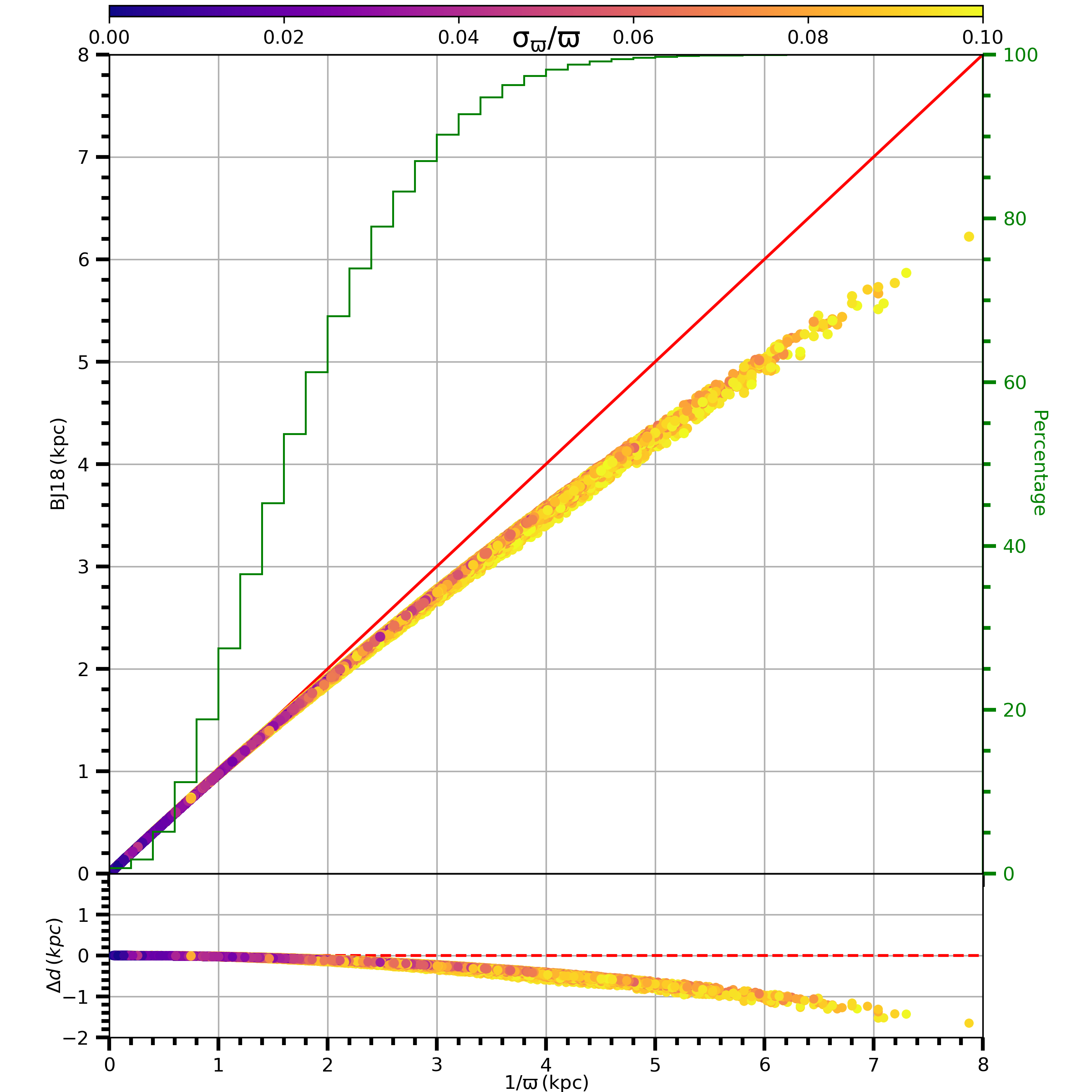}
\caption{Distance comparison between 1/$\varpi$ and BJ18 methods for RGB stars (upper panel). Individual relative parallax errors of RGB stars are represented with colours. Also, on the right y-axis, and on the background the increment of the sample with increasing distances is shown with green step function. The residual distances are shown with $\langle \Delta\rangle$=-0.08 kpc and $\sigma \langle \Delta\rangle$=0.13 kpc. Red dashed line represents the zero line (lower panel).} 
\label{fig:Fig4}
\end {figure}

Stellar distances are obtained using the {\it Gaia} DR2 trigonometric parallaxes of the RGB stars with two methods; the inverse of trigonometric parallaxes (1/$\varpi$) and Bayesian statistics based on the probability analysis \citep[][BJ18]{Bailer-Jones15, Bailer-Jones18}. Comparison between stellar distances with 1/$\varpi$ and BJ18 methods and their residuals are shown in Fig. \ref{fig:Fig4}. Data points are coloured based on the relative parallax errors. The increase in percentages of the RGB stars with the distance is given with the histograms in the upper panel. It is noticed that RGB stars reach up to 8 kpc and 87\% of the sample lies within 3 kpc distances in the Fig. \ref{fig:Fig4}. Distances start to deviate from each other at 1 kpc and this becomes apparent at 2 kpc. Similar comparison was performed at \citet{Plevne20} for their RC sample. The distance difference becomes larger with increasing distance and reaches to 1 kpc at 6 kpc distance in 1/$\varpi$ scale. The mean distance residual is -0.08 kpc and it's dispersion is 0.13 kpc. This implies the distance estimation method is not a critical ingredient up to 2 kpc. 

\begin{figure}
\centering
\includegraphics[width=0.7\textwidth]{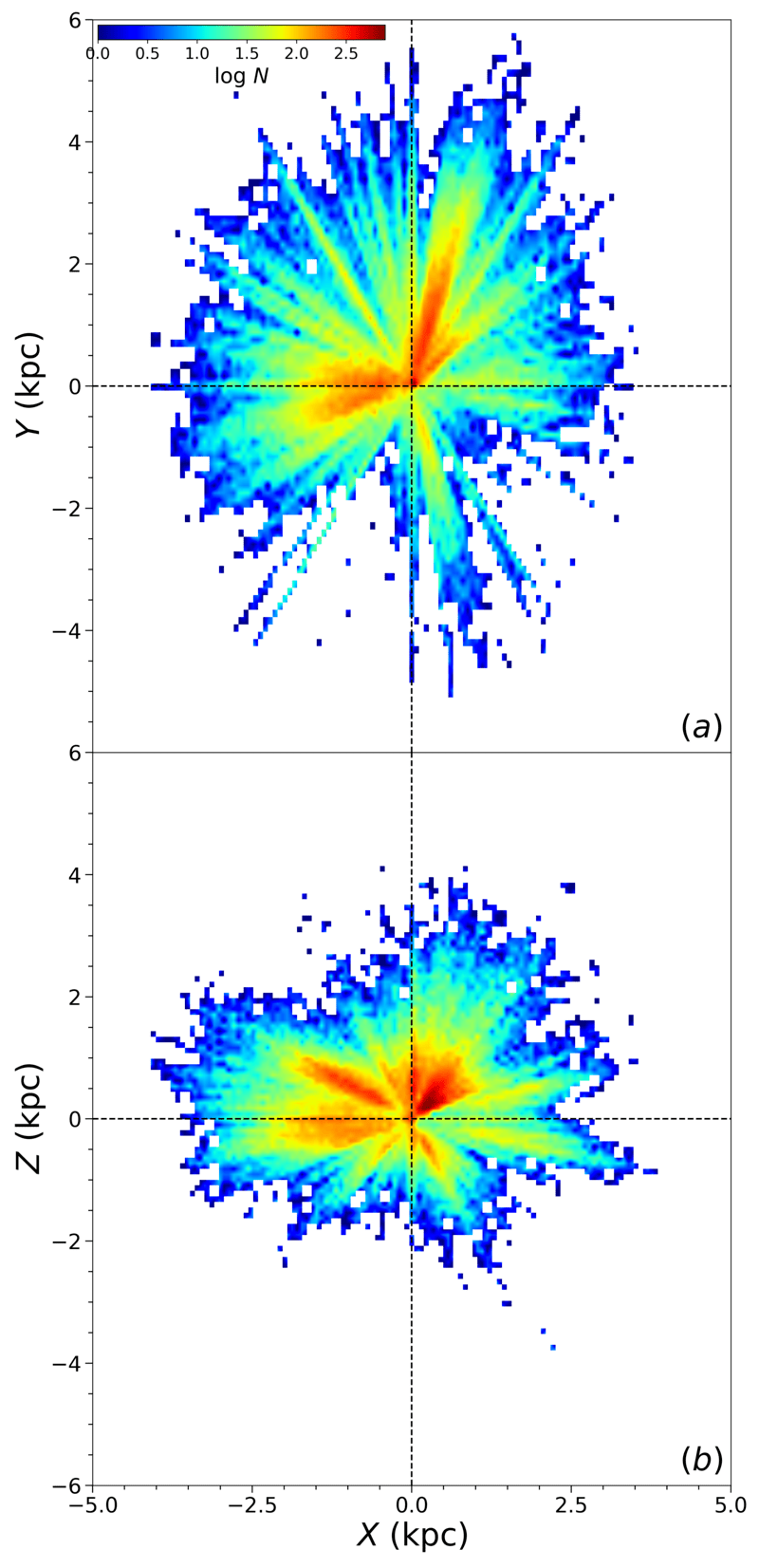}
\caption{Spatial distribution of RGB stars on heliocentric $Y \times X$ (a) and $Z\times X$ (b) planes.} 
\label{fig:Fig5}
\end {figure}

In order to see the spatial distribution of 104,241 RGB stars, the Heliocentric coordinate components ($X$, $Y$, $Z$) are calculated using their Galactic coordinates ($l$, $b$) and distances ($d$). Heliocentric distance distributions of RC stars on $Y \times X$ (a) and $Z \times X$ (b) planes for 1/$\varpi$ methods are given in Fig. \ref{fig:Fig5}. Figure is colour-coded with the logarithmic number density. Median distance to the Sun and median heliocentric distance components with their respective standard errors are 3.24 kpc and ($X$, $Y$, $Z$)=($-0.660\pm0.004$, $+0.440\pm0.004$, $+0.360\pm0.003$) kpc, respectively. 

\subsection{Space Velocity Components}
This study is based on the kinematic properties of RGB stars. The accuracy of the kinematic properties is increased with the quality of astrometric data and with the high resolution and $S/N$ spectroscopic data compiled from known sky surveys. Space velocity components ($U$,$V$,$W$) and their respective uncertainties ($U_{\rm err}$, $V_{\rm err}$, $W_{\rm err}$) of the RGB sample are calculated using the algorithm and matrices of \citet{Johnson87} for the J2000 epoch for the right-handed Galactic reference system. This means that $U$ is positive towards the Galactic centre, $V$ is in the direction of Galactic rotation, and $W$ is positive towards the North Galactic Pole. The total space velocity ($S$) for each star is calculated as $S = \sqrt{U^2 + V^2 + W^2}$.

\begin{figure}
\centering
\includegraphics[width=0.6\columnwidth]{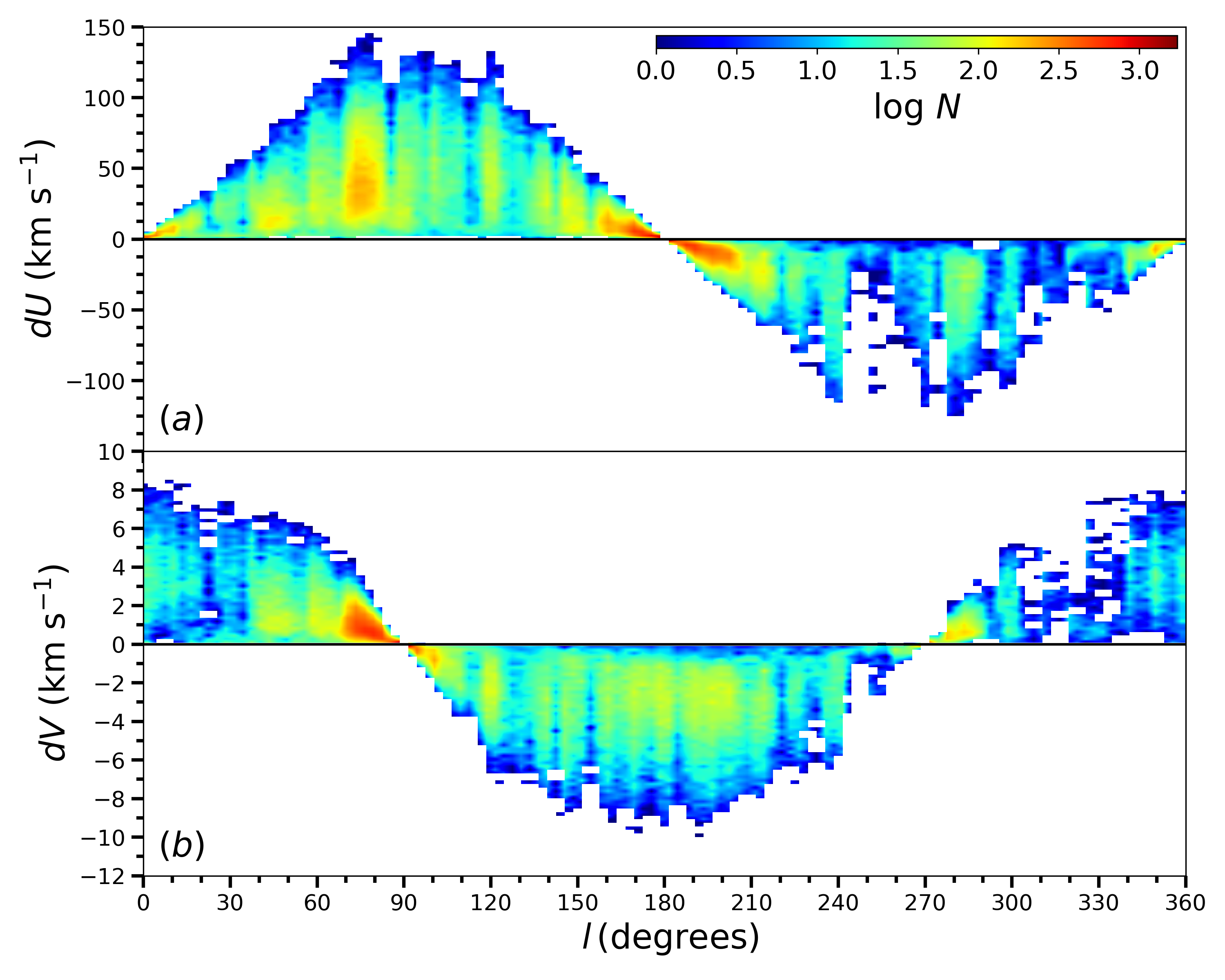}
\caption{Variations of $dU$ (a) and $dV$ (b) differential rotation corrections with Galactic longitude.} 
\label{fig:Fig6}
\end {figure}

Two out of three space velocity components ($U$ and $V$) are directed parallel to the Galactic plane. Since the Galactic disc rotates differentially starting from around 2 kpc to the outer layers, then $U$ and $V$ should be corrected to compensate for the impact on the orbital speeds with these distances from the Galactic centre by applying \citet{Mihalas81}'s differential rotation correction formulae. Variations of $dU$ and $dV$ with respect to the Galactic longitudes are shown in Fig. \ref{fig:Fig6}. Calculation shows the differential rotation correction vary between $-120<dU ({\rm km s}^{-1})< +150$ and $-10<dV ({\rm km s}^{-1})<+10$. The median distance of the RGB sample shows that this correction is significant. After the correction $U$, $V$ and $W$ components are reduced for the local standard of rest by using the $(U, V, W)_{\odot}=(8.83\pm0.24,14.19\pm0.34, 6.57\pm0.21)$ km s$^{-1}$ values of \cite{Coskunoglu11}.

\begin{figure}
\centering
\includegraphics[width=0.5\columnwidth]{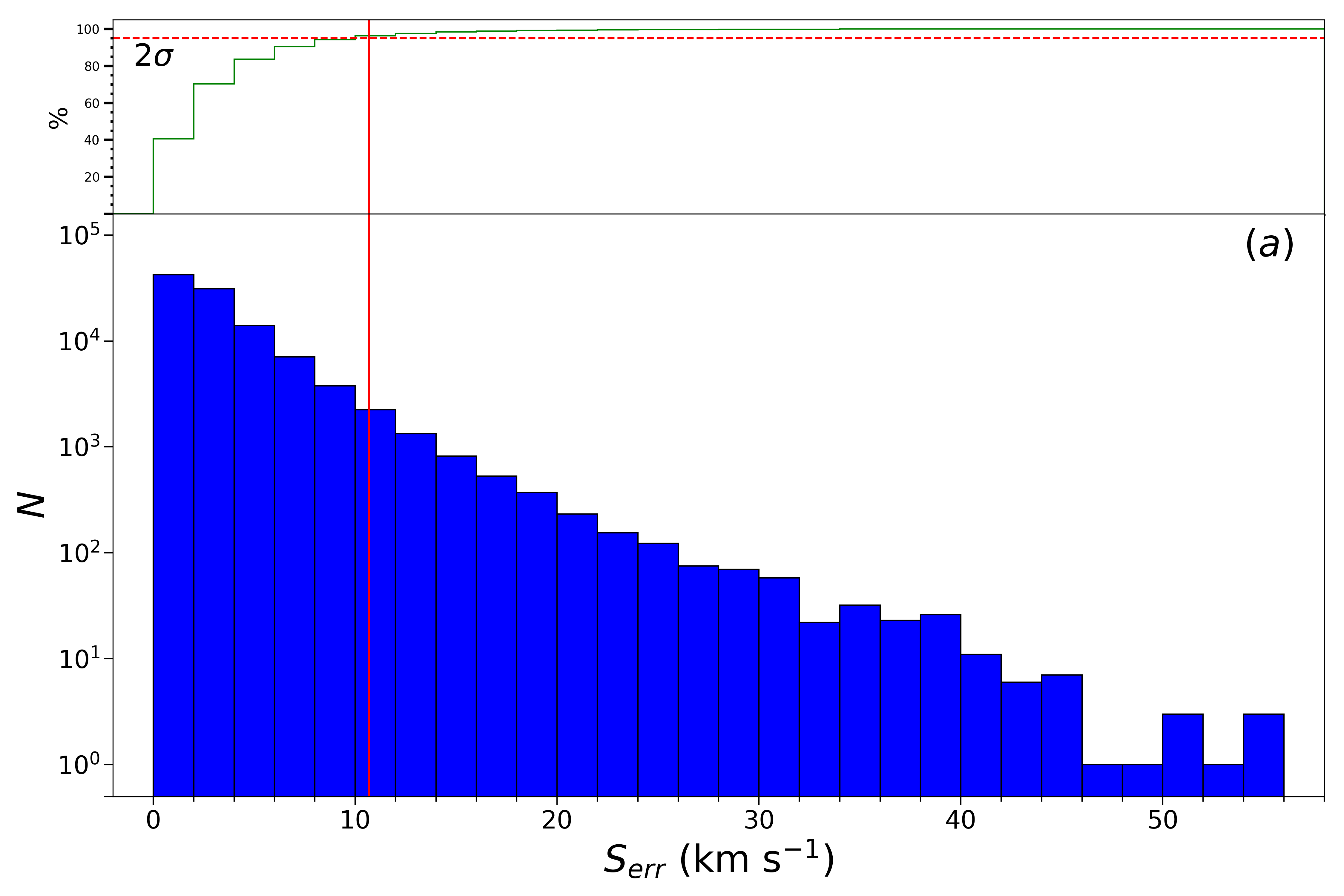}
\includegraphics[width=0.5\columnwidth]{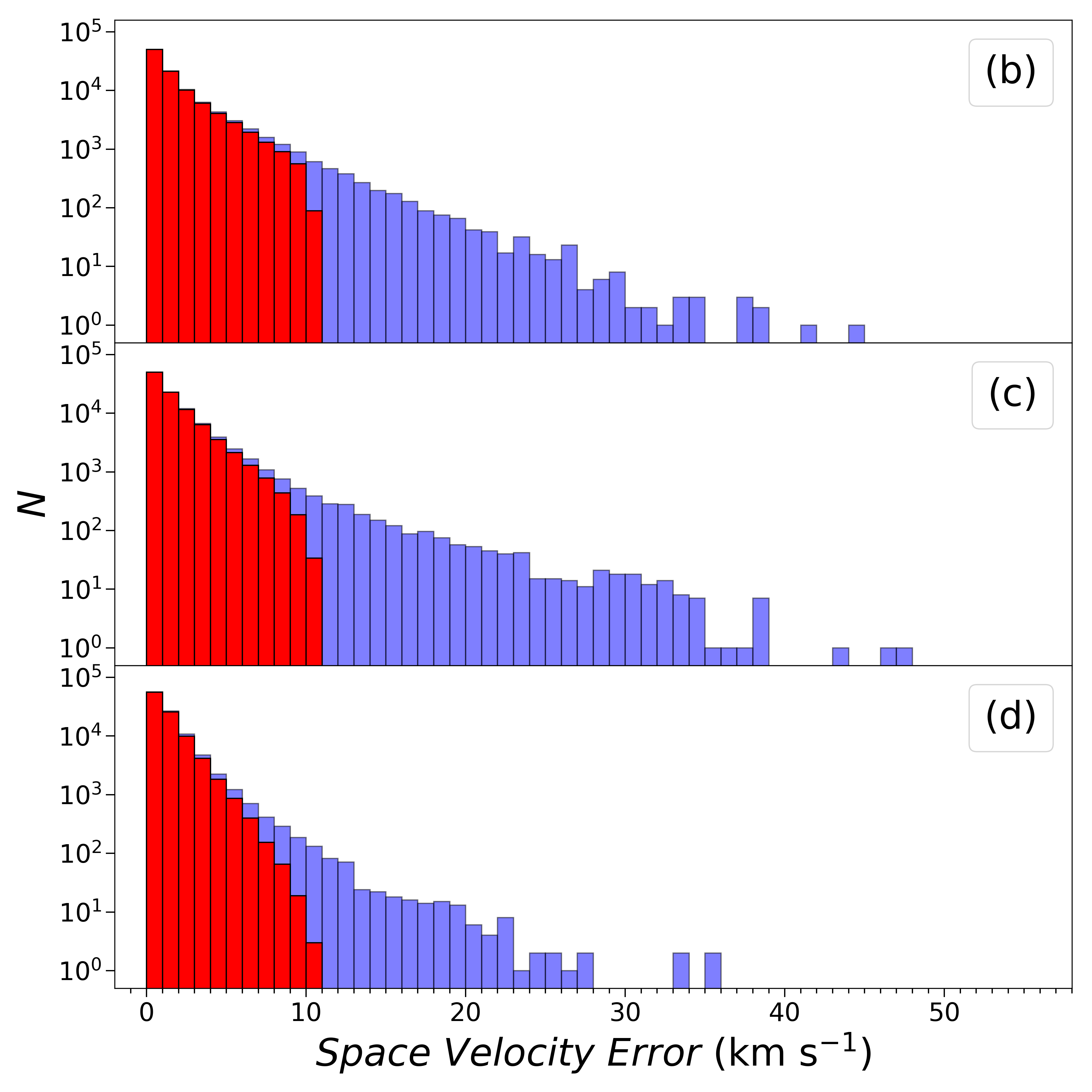}
\caption{$S_{\rm err}$ histogram of 104,241 RGB stars and cumulative change of stellar number (a). Red solid line represents the 2$\sigma$ limit on $S_{\rm err}$ axis, while red dashed line represents the corresponding number of stars to 2$\sigma$. Error histograms of space velocity components, i.e. $U_{\rm err}$ (b), $V_{\rm err}$ (c), and $W_{\rm err}$ (d), before (purple) and after (red) the $S_{\rm err}$=10.68 km s$^{-1}$ cut-off.} 
\label{fig:Fig7}
\end {figure}

Propagating the uncertainties in radial velocity, trigonometric parallax and proper motion components of RGB stars using \citet{Johnson87}’s algorithm gives the uncertainty on each space velocity component. Then, the total space velocity error ($S_{\rm err}$) is obtained from these uncertainties for each star from $S_{\rm err}=\sqrt{U_{\rm err}^2 + V_{\rm err}^2 + W_{\rm err}^2}$ relation. $S_{\rm err}$ distribution and the cumulative variation of the percentile number of stars and the error histograms of RGB sample are shown in Fig. \ref{fig:Fig7}. In panel (a), the blue histogram shows the raw distribution of the total space velocity, which reaches up to $S_{\rm err}=60$ km s$^{-1}$, while the red line represents the 2$\sigma$ value. Step function shows the number of RGB stars with the cumulative increase of RGB stars with $S_{\rm err}$. 

Accuracy of the kinematic parameters is enhanced by applying a final constraint on the $S_{\rm err}$ distribution. Based on the 2$\sigma$ value of the distribution a cut-off point is defined as 10.68 km s$^{-1}$. Eliminating the stars beyond this cut-off point provides the most sensitive kinematic RGB sample, which is composed of 99,029 stars. Median space velocity errors of this sample are $(U_{\rm err}, V_{\rm err}, W_{\rm err}) = (1.06\pm0.003,1.06\pm0.003,0.90\pm0.003)$ km s$^{-1}$. In panels of Fig. \ref{fig:Fig7}, distributions of the individual space velocity component errors ($U_{\rm err}$, $V_{\rm err}$, $W_{\rm err}$) are shown before (light-blue histograms) and after (red histograms) the cut-off.

\begin{figure}
\centering
\includegraphics[width=0.55\columnwidth]{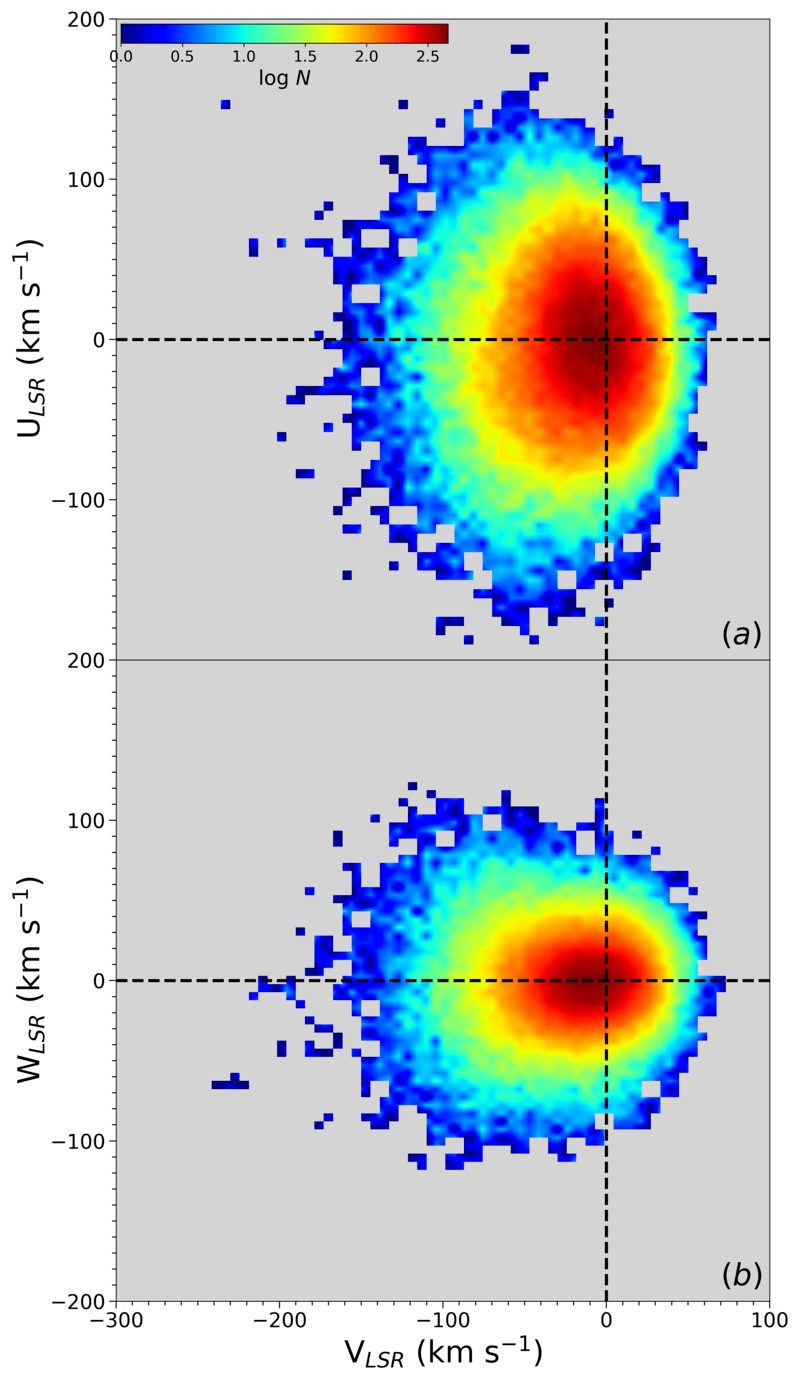}
\caption{Distribution of velocity components of RGB stars on to the Galactic planes: $U_{\rm LSR} \times V_{\rm LSR}$ (a) and $W_{\rm LSR} \times V_{\rm LSR}$ (b). RGB stars are colour coded with logarithmic stellar number density. Black dashed-lines in both panels show the centroid of the LSR as 0 km s$^{-1}$. } 
\label{fig:Fig8}
\end {figure} 

After correcting the space velocity components for the differential rotation and LSR, and constraining the RGB sample for $S_{\rm err}$, the distribution of the space velocity components of 99,029 RGB stars on the $U_{\rm LSR} \times V_{\rm LSR}$ and $W_{\rm LSR} \times V_{\rm LSR}$ planes are shown in Fig. \ref{fig:Fig8}. The figure is colour coded for logarithmic number density of RGB stars where space velocity components vary $-200\leq U_{\rm LSR}~({\rm km~s^{-1}}) \leq200$, $-240\leq V_{\rm LSR}~ ({\rm km~s^{-1}}) \leq70$, and $-120\leq W_{\rm LSR}~({\rm km~s^{-1}}) \leq120$ intervals. According to the positions of RGB stars on these planes, stars with small space velocity errors are moving at Solar like velocities. By doing so the velocity dispersion calculations are precise enough to use in age calculations for an AMR investigation. Dispersion of space velocity components ($\sigma_{\rm U}, \sigma_{\rm V}, \sigma_{\rm W}$) are calculated using Eq. \ref{eq: dispersion_formula} below: 

\begin{eqnarray}
\sigma_{\rm U} = \sqrt{\sum_{i=1}^{N}\frac{U_i^2}{N} - \left(\sum_{i=1}^{N}\frac{U}{N}\right)^2 - \sum_{i=1}^{N}\frac{U_{{\rm err_i}}^2}{N}}, \nonumber \\
\sigma_{\rm V} = \sqrt{\sum_{i=1}^{N}\frac{V_i^2}{N} - \left(\sum_{i=1}^{N}\frac{V}{N}\right)^2 - \sum_{i=1}^{N}\frac{V_{{\rm err_i}}^2}{N}}, \nonumber \\
\sigma_{\rm W} = \sqrt{\sum_{i=1}^{N}\frac{W_i^2}{N} - \left(\sum_{i=1}^{N}\frac{W}{N}\right)^2 - \sum_{i=1}^{N}\frac{W_{{\rm err_i}}^2}{N}}.
 \label{eq: dispersion_formula}
\end{eqnarray}
here $U_{\rm i}$, $V_{\rm i}$, and $W_{\rm i}$ and $N$, represent space velocity components for each star and total number of stars in the stellar sub-sample, respectively.

\begin{figure}
\centering 
\includegraphics[width=0.6\columnwidth]{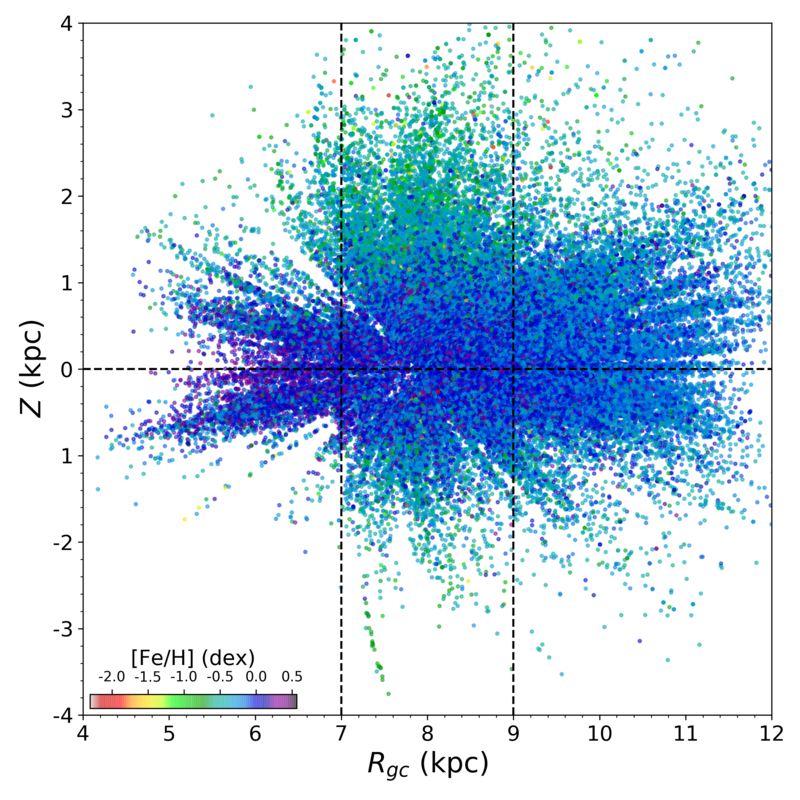}
\caption{Distribution of RGB stars coloured by metallicity on the $Z \times R_{\rm gc}$ plane. Vertical dashed lines represent the borders of the solar cylinder ($7\leq R_{\rm gc} {\rm (kpc)}\leq 9$) and horizontal dashed lines shows Galactic plane.} 
\label{fig:Fig9}
\end {figure}

AMR is investigated on the current Galactocentric radial distance ($R_{\rm gc}$) from the Galactic centre and current vertical distance ($Z$) from the Galactic plane. $R_{\rm gc}$ values of RGB stars are calculated using the formula derived from the geometry \citep[see][]{Tuncel19}. The distribution of our RGB sample on $Z\times R_{\rm gc}$ plane is shown in Fig. \ref{fig:Fig9}, where the stars are colour coded by their metallicity. Galactic plane (at $Z=0$) and the Solar cylinder ($7\leq R_{\rm gc}~{\rm (kpc)}\leq9$) borders are presented with the black dashed lines in Fig. \ref{fig:Fig9}. In this study, the AMR investigations are focused on the Solar cylinder, where 43,592 RGB stars lie inside these borders. This is the final sample used in AMR analysis. It is noticed that there are three times more stars lie at $Z>0$ kpc (32,187 stars) than $Z\leq0$ kpc (11,405 stars). 

\section{Methods}
This study aims to investigate the age and metallicity trends within the Galactic disc. There are strong evidence that the Milky Way disc is composed of overlapping but distinct in nature (age, chemistry and kinematics) sub-components, i.e., thin and thick discs discussed by earlier studies such as \citet{GW85, Gratton96, Fuhrmann98, Chiba00,Navarro11, Karaali19, Plevne20}. In the thin and thick discs, iron and $\alpha$-element abundance ratios change towards the radial direction from the Galactic centre or in the vertical direction from the Galactic plane \citep{Bilir06, Bilir08, Bilir12, Ak07a, Ak07b, Cabrera07, Yaz10}. Distribution of stars on the [$\alpha$/Fe]$\times$[Fe/H] plane is a reflection of the chemical evolution of the Milky Way. Investigations of the chemical evolution of the Galactic disc have shown that the disc has undergone at least two different periods of the formation according to the renowned two infall model of \citet{CMG97}. The stars born during these formation periods can be separated on [$\alpha$/Fe]$\times$[Fe/H] plane. It is also known that the chemical abundances plotted on this plane can be used to disentangle the components of the Galactic disc in age \citep{Wyse88}.

\subsection{Classification of Galactic Populations}
\begin{figure}
\centering
\includegraphics[width=\columnwidth]{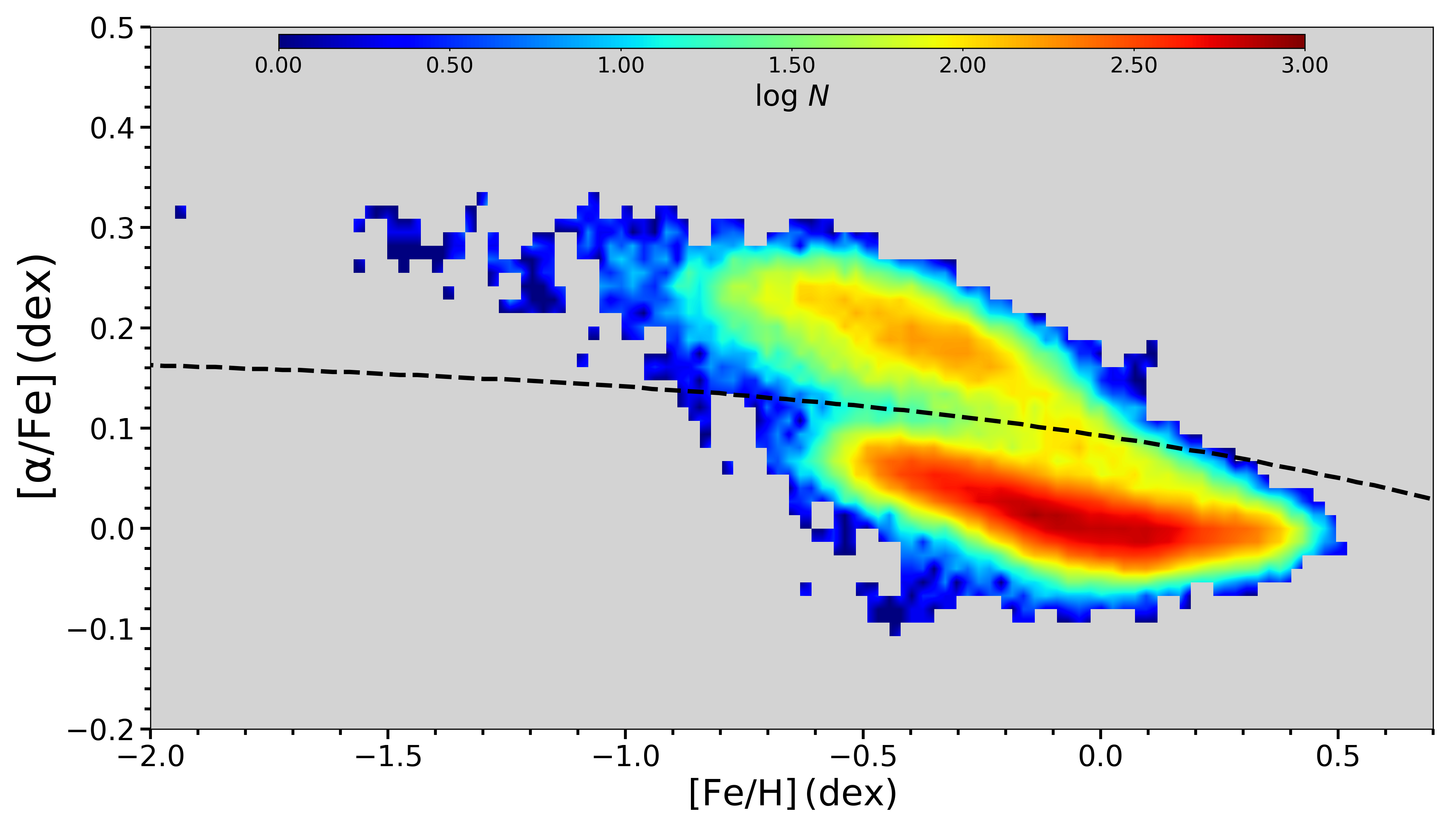}
\caption{Logarithmic number density distribution of 43,592 RGB stars on [$\alpha$/Fe]$\times$[Fe/H] plane. Black dashed line presents the decision boundary obtained with Gaussian mixture model.}
\label{fig:Fig10}
\end {figure}

 To determine Galactic populations, this study made use of Gaussian Mixture Model (GMM), an unsupervised machine learning algorithm which classifies the data by fitting preferred number of Gaussian planes. We applied GMMs to the [$\alpha$/Fe]$\times$[Fe/H] plane of RGB stars. Then, it calculates the probabilities of each Gaussian plane. GMM is applied using {\it sklearn} version 0.19.1 \citep{Pedregosa11}. For more details about the procedure applied in this study, see \citet{Plevne20}. The GMM procedure is applied to 43,592 RGB stars that have [$\alpha$/Fe] and [Fe/H] abundance ratios. Iron abundances range $\rm -2\leq{[Fe/H]~(dex)}\leq 0.5$, while $\alpha$ abundances range $\rm -0.2\leq{[\alpha/Fe]~(dex)}\leq 0.5$. Results of GMM on the [$\alpha$/Fe]$\times$[Fe/H] plane is shown in Fig. \ref{fig:Fig10}. The figure is colour-coded for logarithmic number density, and the black dashed line is the decision boundary line that allows categorising RGB stars to chemical Galactic populations such as low-[$\alpha$/Fe] or high-[$\alpha$/Fe] population objects (hereafter low-$\alpha$ and high-$\alpha$, respectively). GMM classification is based on the decision boundary line. According to this, 36,204 stars are classiefied as low-$\alpha$, i.e. chemical thin disc, while 7,388 stars are classified as high-$\alpha$, i.e. chemical thick disc.

\subsection{Kinematic Age}
Space velocities of stars change systematically causing a departure behaviour from the Galactic plane as a function of their age and radial distance from the Galactic centre. If these stars are involved with any stellar group, this also increases the total space velocity dispersion and the mean value of the group age. This fact is provided as a base for a new method to obtain the stellar age. As the stars move away from the Galactic plane they also move to a less dense medium, meanwhile their interaction with surrounding objects increases space velocity $v$ and decreases diffusion co-efficent \citep{Spitzer53, Chandrasekhar60, Wielen77}. 

Kinematic age estimation is based on the dispersion of true space motions and requires a complete set of space velocity components. Age is calculated from the velocity dispersion of a statistically significant RGB sample using the equation given in \citet{Cox00}, which is the improved version of \citet{Wielen77}'s study.

\begin{equation}
\sigma_{\nu}^{3}(\tau)=\sigma_{\nu,\tau=0}^{3}+\frac{3}{2}\alpha_{V}\delta_{2}T_{\delta}
\Biggl[\exp\Biggl(\frac{\tau}{T_{\delta}}\Biggl)-1\Biggr],
\end{equation}
where, $\sigma_{\nu,\tau=0}$ is the velocity dispersion at zero age, which is usually taken as 10 km s$^{-1}$ \citep{Cox00}, $T_{\delta}$ is a time scale ($5\times10^{9}$ yr), $\delta_{2}$ is a diffusion coefficient ($3.7\times10^{-6}$ (km~s$^{-1})^{3}$ yr), and $\alpha_{V}$ is a parameter describing the rotation curve ($\approx2.95$). $\sigma_{\nu}(\tau)$ is the total velocity dispersion of the group of RGB. $\tau$ is the kinematic age of the group. The total dispersion of space velocity vectors is connected to the dispersion of the velocity components as

\begin{equation}
\sigma_{\nu}=\sqrt{\sigma_{U{_{\rm LSR}}}^{2}+\sigma_{V{_{\rm LSR}}}^{2}+\sigma_{W{_{ \rm LSR}}}^{2}}.
\end{equation}
After computing $\sigma_{\nu}$ from the dispersion of velocity components, it is used in Eq. (2) for computing kinematic age ($\tau$).

\section{Results and Discussion}
First of all, our sample stars lie within the Solar cylinder ($7\leq R_{\rm gc} (\rm kpc)\leq9$) that extends up to $|Z|=4$ kpc, effectively. RGB stars are examined by their kinematics and their chemical properties. GMM model divides the sample into chemical populations which are called as low-$\alpha$ (thin disc) ve high-$\alpha$ (thick disc/halo). Each of the chemical disks separated into the $Z$ distance intervals.For each $Z$ intervals following parameters are calculated: (i)the space velocity components and their respective dispersions, (ii) total space velocity dispersions, (iii) kinematic age and (iv) metallicities. Total velocity dispersions are calculated with the Eq. (3), while kinematic ages are calculated with Eq. (2). The space velocity components and the metallicities are determined by the histograms and mean values calculated from the fitted Gaussians. The kinematic and chemical results of the low-$\alpha$ (upper panel) and high-$\alpha$ (lower panel) populations in different $Z$ distance intervals are listed in Table 1. 
 
\subsection{AMR for the Low-$\alpha$ Population}
\begin{figure*}
\centering
\includegraphics[width=\textwidth]{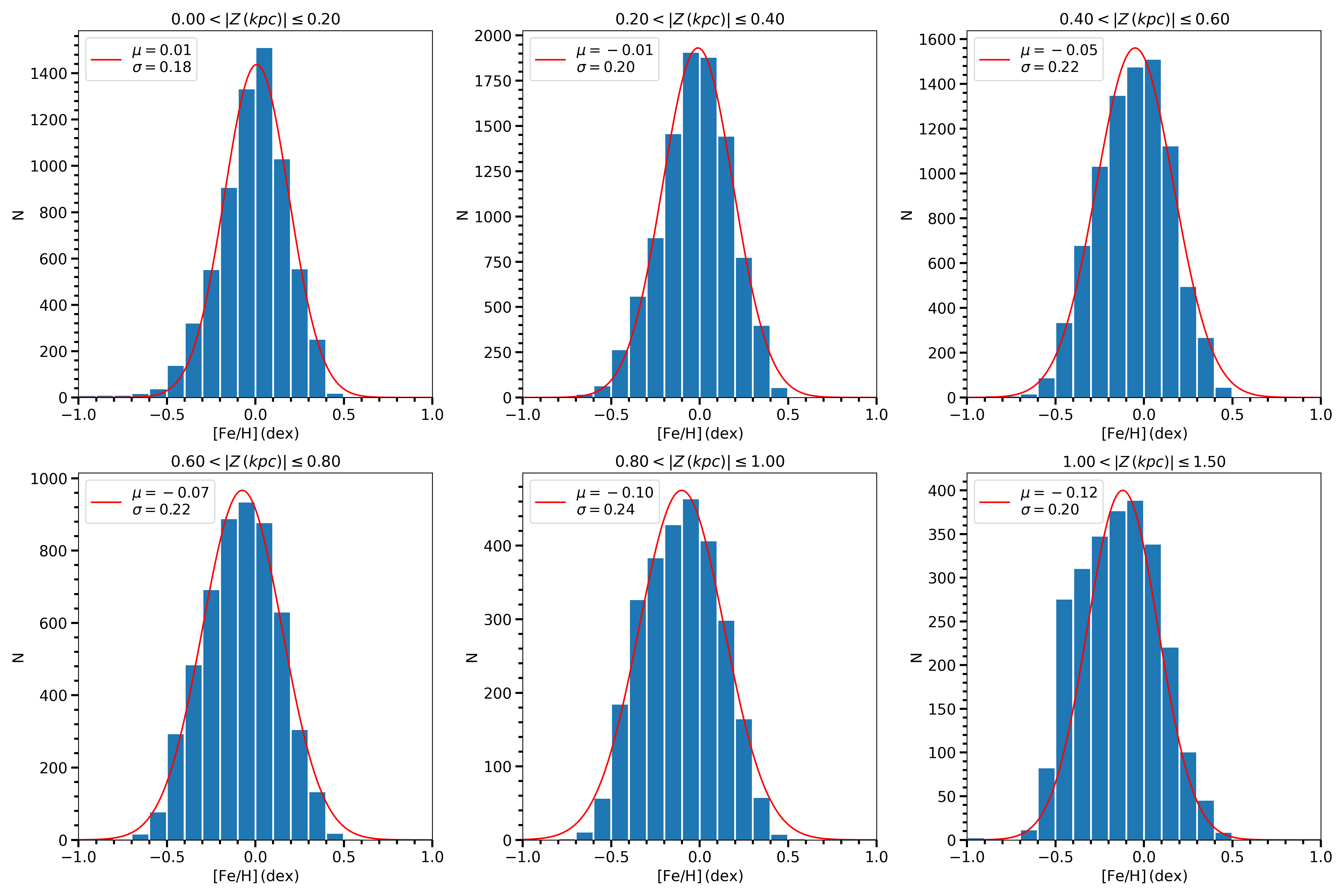}
\caption{Metallicity histograms for low-$\alpha$ population for different $Z$ distance intervals. Blue bars are the metallicity distributions and red solid lines are the Gaussian fits to distributions in each distance interval.}
\label{fig:Fig11}
\end {figure*}

\begin{figure}[th]
\centering
\includegraphics[width=0.70\textwidth]{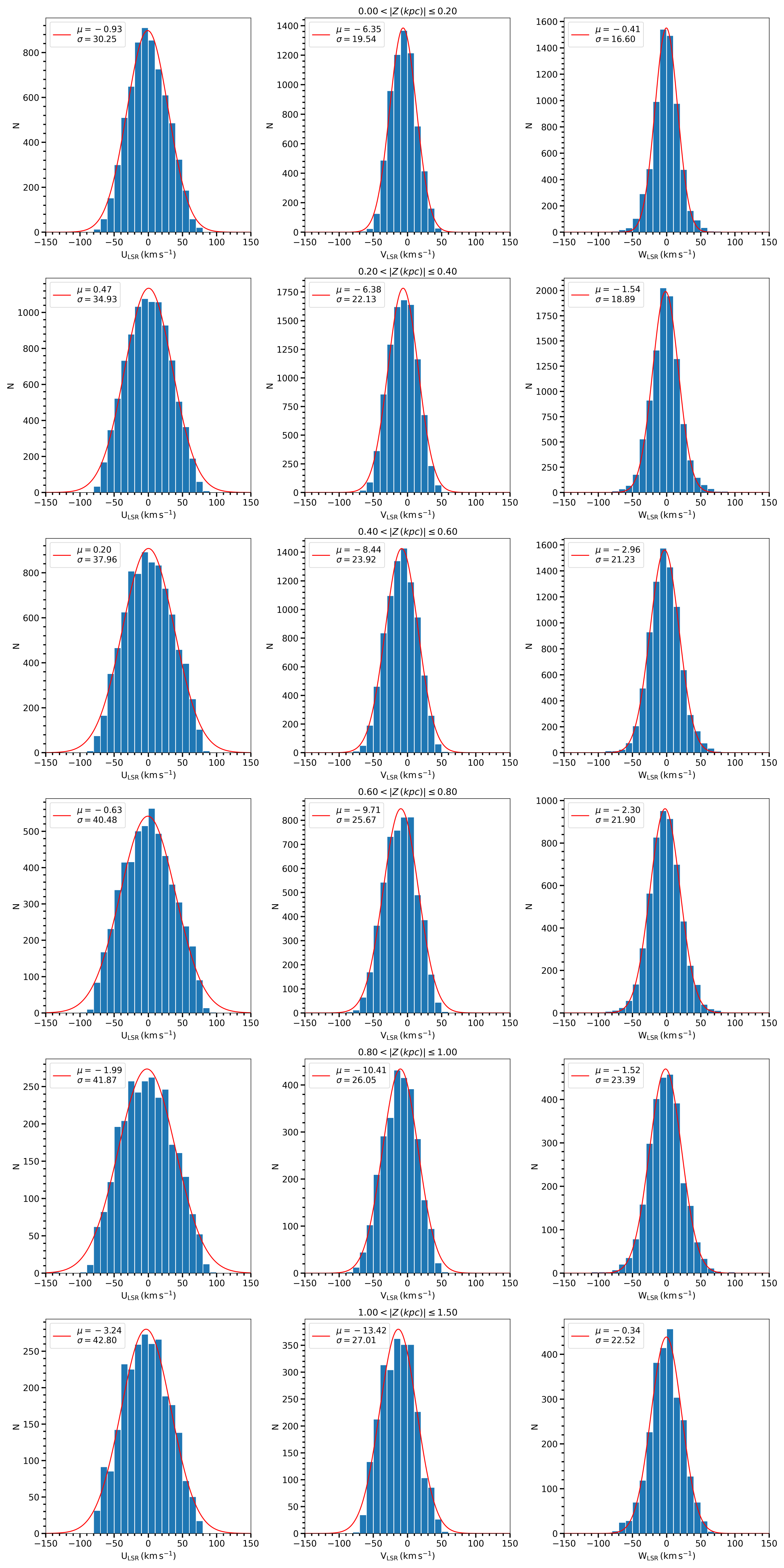}
\caption{Velocity histograms for low-$\alpha$ population for different $Z$ distance intervals. Blue bars are the velocity distributions and red solid lines are the Gaussian fits to distributions in each distance interval.}
\label{fig:Fig12}
\end {figure}

The RGB stars are separated into different $Z$ intervals to probe the variation of the kinematic age with metallicity. Low-$\alpha$ RGB stars are separated with 0.2 kpc steps up to 1 kpc. Also, there are two more distance intervals that are $1\leq |Z|\leq1.5$ and $|Z|>1.5$ kpc. $U$, $V$ and $W$ space velocity components and their dispersions are calculated for stars in these distance intervals. Metallicity and space velocity component distributions for each distance intervals and plotted in Fig. \ref{fig:Fig11} and Fig. \ref{fig:Fig12}, respectively. These distributions are presented with blue bars and red solid lines are the Gaussians fitted to the histograms. Mean values ($\mu$) and standard deviations ($\sigma$) are given in each panel. Metallicity follow a Gaussian trend in all panels. It decreases from -0.010$\pm$0.002 dex to -0.120$\pm$0.004 dex as a function of increasing $|Z|$ distances. According to Table 1, kinematic age and total space velocity dispersion increase, while metallicity and $V_{\rm LSR}$ velocity decrease with $|Z|$. Kinematic age of RGB stars vary between 4.37$\pm$0.28 to 6.57$\pm$0.26 Gyr. There is therefore a $\sim$ 2.2 Gyr age difference among the low-$\alpha$ RGB population. Each space velocity component varies in a narrow range when all $|Z|$ intervals considered in low-$\alpha$ sub-samples. The mean value of $V_{\rm LSR}$ is about -9.12 km s$^{-1}$, which is in agreement with the mean $V$=-8 km s$^{-1}$ value of the \textit{Gaia} DR1 TGAS APOGEE sample of \citet{Allende16}. Moreover, the mean value of $W_{\rm LSR}$=-1.51 km s$^{-1}$ is also closer to the $W=0$ km s$^{-1}$ velocity of the same study. However, there is a prominent difference between mean $U_{\rm LSR}$ value of the two samples. In \citet{Allende16}'s study, the space velocity dispersions of each velocity have been investigated for the stars within the Solar cylinder (i.e. $7.5\leq R_{\rm gc} (\rm kpc)\leq9.5$). They found velocity dispersions of low-$\alpha$ stars at $|Z|<0.5$ kpc as $(\sigma_{\rm U}, \sigma_{\rm V}, \sigma_{\rm W})=(28,22,18)$ km s$^{-1}$, In our study, the mean velocity dispersions of each space velocity component found as $(\sigma_{\rm U}, \sigma_{\rm V}, \sigma_{\rm W})=(38,24,21)$ km s$^{-1}$.  Both studies are in agreement.

\subsection{AMR for the High-$\mathrm{\alpha}$ Population}
\begin{figure}
\centering
\includegraphics[width=\columnwidth]{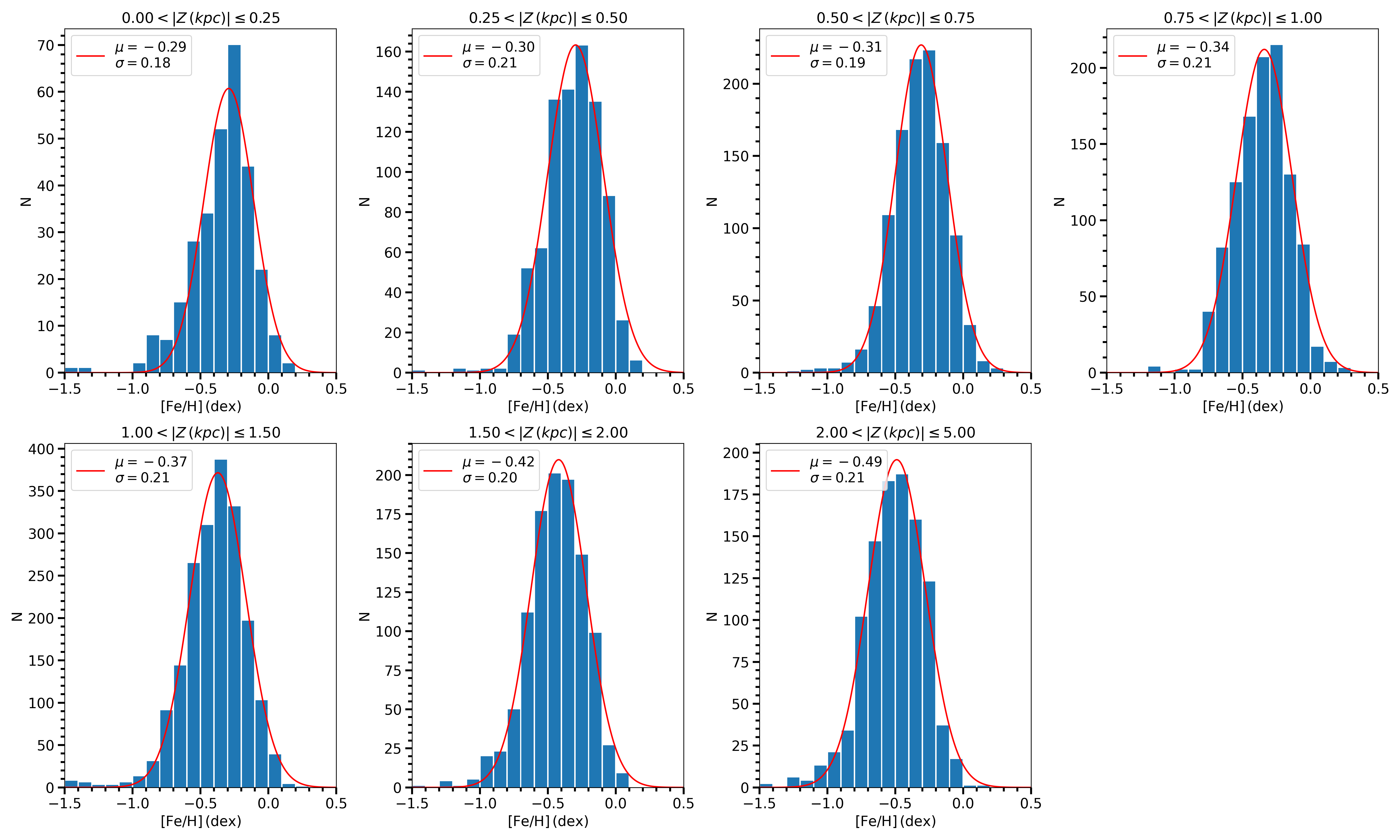}
\caption{Metallicity histograms for high-$\alpha$ population for different $Z$ distance intervals. Blue bars are the metallicity distributions and red solid lines are the Gaussian fits to distributions in each distance interval.}
\label{fig:Fig13}
\end {figure}

The same analysis routine is applied to the high-$\alpha$ population for seven subsequent different distance intervals according to Table 1. These intervals increase with 0.25 kpc steps up to 1 kpc, then there are two more additional distance intervals, i.e. $1.5<Z{\rm (kpc)}\leq2$, and $2<Z{\rm (kpc)}\leq5$ (see Fig. \ref{fig:Fig13}). According to the table, kinematic age and total space velocity dispersion increase, while metallicity and $V_{\rm LSR}$ velocity decrease with $Z$ as these are the expected trends for the Galactic disc \citep[i.e][]{Boeche13}. Kinematic age of the RGB stars is 6.62$\pm$0.18 Gyr for the high-$\alpha$ stars that are embedded in the Galactic plane ($|Z|\leq0.2$ kpc), and age increases to 8.73$\pm$0.29 Gyr for stars with $|Z|>2$ kpc. There is a $\sim$ 2.1 Gyr age difference among the high-$\alpha$ RGB population. Iron abundance ratios are in agreement with the relatively metal-rich part of the metallicity distribution of the Galactic high-$\alpha$ population, which vary between -0.29$\pm$0.010 to -0.49$\pm$0.006 dex (Fig. \ref{fig:Fig13}).

\begin{landscape}
\begin{table*}
\setlength{\tabcolsep}{2pt}
\renewcommand{\arraystretch}{1.5}
\centering%

\caption{Kinematic properties for the low and high-$\alpha$ populations. Perpendicular distance intervals from the Galactic plane ($|Z_1-Z_2|$), number of stars ($N$) in each distance interval, LSR corrected space velocity components ($U_{\rm LSR}, V_{\rm LSR}, W_{\rm LSR}$) and their dispersions ($\sigma_{\rm U}, \sigma_{\rm V}, \sigma_{\rm W}$), total space velocity dispersion ($\sigma_{\rm tot}$), kinematic age ($\tau$) and the metallicity ([Fe/H]) are given in columns, respectively.}
\begin{tabular}{ccccccccccc}%
\hline
\hline
\multicolumn{11}{c}{low-[$\alpha$/Fe] population}\\
\hline
 $|Z_1 - Z_2|$ & $N$ & $U_{\rm LSR}$ & $V_{\rm LSR}$ & $W_{\rm LSR}$ & $\sigma_{\rm U}$ & $\sigma_{\rm V}$&  $\sigma_{\rm W}$ &  $\sigma_{\rm tot}$ & $\tau$ & [Fe/H] \\
(kpc) & (stars) & (km s$^{-1}$) & (km s$^{-1}$) & (km s$^{-1}$) & (km s$^{-1}$) & (km s$^{-1}$) & (km s$^{-1}$) & (km s$^{-1}$) & (Gyr) & (dex)  \\
\hline

0.00-0.20 & 6691 & $-0.93 \pm 1.06$ & $-6.35 \pm 0.50$ & $-0.41 \pm 0.64$ & $30.25 \pm 1.47$ & $19.54 \pm 0.60$ & $16.60 \pm 0.58$ & $39.65 \pm 1.69$ & $4.37 \pm 0.28$ & $-0.01 \pm 0.002$ \\
0.20-0.40 & 9687 & $+0.47 \pm 1.37$ & $-6.38 \pm 0.58$ & $-1.54 \pm 0.72$ & $34.93 \pm 1.60$ & $22.13 \pm 0.63$ & $18.89 \pm 0.63$ & $45.46 \pm 1.83$ & $5.28 \pm 0.26$ & $-0.01 \pm 0.002$ \\
0.40-0.60 & 8401 & $+0.20 \pm 1.83$ & $-8.44 \pm 0.72$ & $-2.96 \pm 0.84$ & $37.96 \pm 2.00$ & $23.92 \pm 0.66$ & $21.23 \pm 0.75$ & $49.64 \pm 2.23$ & $5.85 \pm 0.30$ & $-0.05 \pm 0.002$ \\
0.60-0.80 & 5341 & $-0.63 \pm 2.08$ & $-9.71 \pm 0.82$ & $-2.30 \pm 0.84$ & $40.48 \pm 2.15$ & $25.67 \pm 0.72$ & $21.90 \pm 0.78$ & $52.70 \pm 2.40$ & $6.25 \pm 0.30$ & $-0.07 \pm 0.003$ \\
0.80-1.00 & 2785 & $-1.99 \pm 2.00$ & $-10.41 \pm 0.98$ & $-1.52 \pm 0.83$ & $41.87 \pm 2.02$ & $26.05 \pm 0.84$ & $23.39 \pm 0.76$ & $54.58 \pm 2.31$ & $6.48 \pm 0.28$ & $-0.10 \pm 0.005$ \\
1.00-1.50 & 2505 & $-3.24 \pm 2.04$ & $-13.42 \pm 1.19$ & $-0.34 \pm 0.80$ & $42.80 \pm 1.88$ & $27.01 \pm 0.94$ & $22.52 \pm 0.71$ & $55.39 \pm 2.22$ & $6.57 \pm 0.26$ & $-0.12 \pm 0.004$ \\
$>1.50$ & 794 & --- & --- & --- & --- & --- & --- & --- & --- & --- \\
\hline
\multicolumn{11}{c}{high-[$\alpha$/Fe] population}\\
\hline
0.00-0.25 & 294 & $+0.74 \pm 0.92$ & $-13.90 \pm 0.54$ & $-2.08 \pm 0.84$ & $34.93 \pm 1.21$ & $23.86 \pm 0.63$ & $36.42 \pm 0.74$ & $55.82 \pm 1.55$ & $6.62 \pm 0.18$ & $-0.29 \pm 0.010$ \\
0.25-0.50 & 839 & $-2.24 \pm 1.58$ & $-19.06 \pm 0.72$ & $-3.17 \pm 0.98$ & $38.34 \pm 1.78$ & $24.54 \pm 0.67$ & $36.40 \pm 0.89$ & $58.28 \pm 2.09$ & $6.91 \pm 0.24$ & $-0.30 \pm 0.007$ \\
0.50-0.75 & 1093 & $+1.47 \pm 1.91$ & $-22.32 \pm 0.92$ & $-1.57 \pm 1.04$ & $39.82 \pm 2.02$ & $27.71 \pm 0.79$ & $37.27 \pm 0.99$ & $61.18 \pm 2.38$ & $7.22 \pm 0.25$ & $-0.31 \pm 0.006$ \\
0.75-1.00 & 1086 & $+0.09 \pm 1.97$ & $-25.92 \pm 1.19$ & $-0.73 \pm 0.97$ & $43.09 \pm 1.98$ & $28.01 \pm 0.98$ & $36.57 \pm 0.91$ & $63.08 \pm 2.39$ & $7.42 \pm 0.25$ & $-0.34 \pm 0.006$ \\
1.00-1.50 & 1952 & $-3.03 \pm 2.21$ & $-33.31 \pm 1.68$ & $-0.85 \pm 1.09$ & $44.18 \pm 1.90$ & $28.83 \pm 1.28$ & $38.09 \pm 1.05$ & $65.07 \pm 2.52$ & $7.63 \pm 0.25$ & $-0.37 \pm 0.005$ \\
1.50-2.00 & 1078 & $+1.83 \pm 2.80$ & $-39.99 \pm 2.38$ & $-3.46 \pm 1.26$ & $44.53 \pm 2.19$ & $28.00 \pm 1.61$ & $41.96 \pm 1.18$ & $67.29 \pm 2.97$ & $7.85 \pm 0.29$ & $-0.42 \pm 0.006$ \\
$>2.00$ & 1046 & $-7.31 \pm 3.62$ & $-42.33 \pm 2.99$ & $-6.06 \pm 1.48$ & $51.46 \pm 2.45$ & $33.49 \pm 2.02$ & $46.48 \pm 1.28$ & $77.01 \pm 3.43$ & $8.73 \pm 0.29$ & $-0.49 \pm 0.006$ \\
\hline
\hline
\end{tabular}%
\end{table*}
\end{landscape}

\begin{figure}
\centering
\includegraphics[width=\columnwidth]{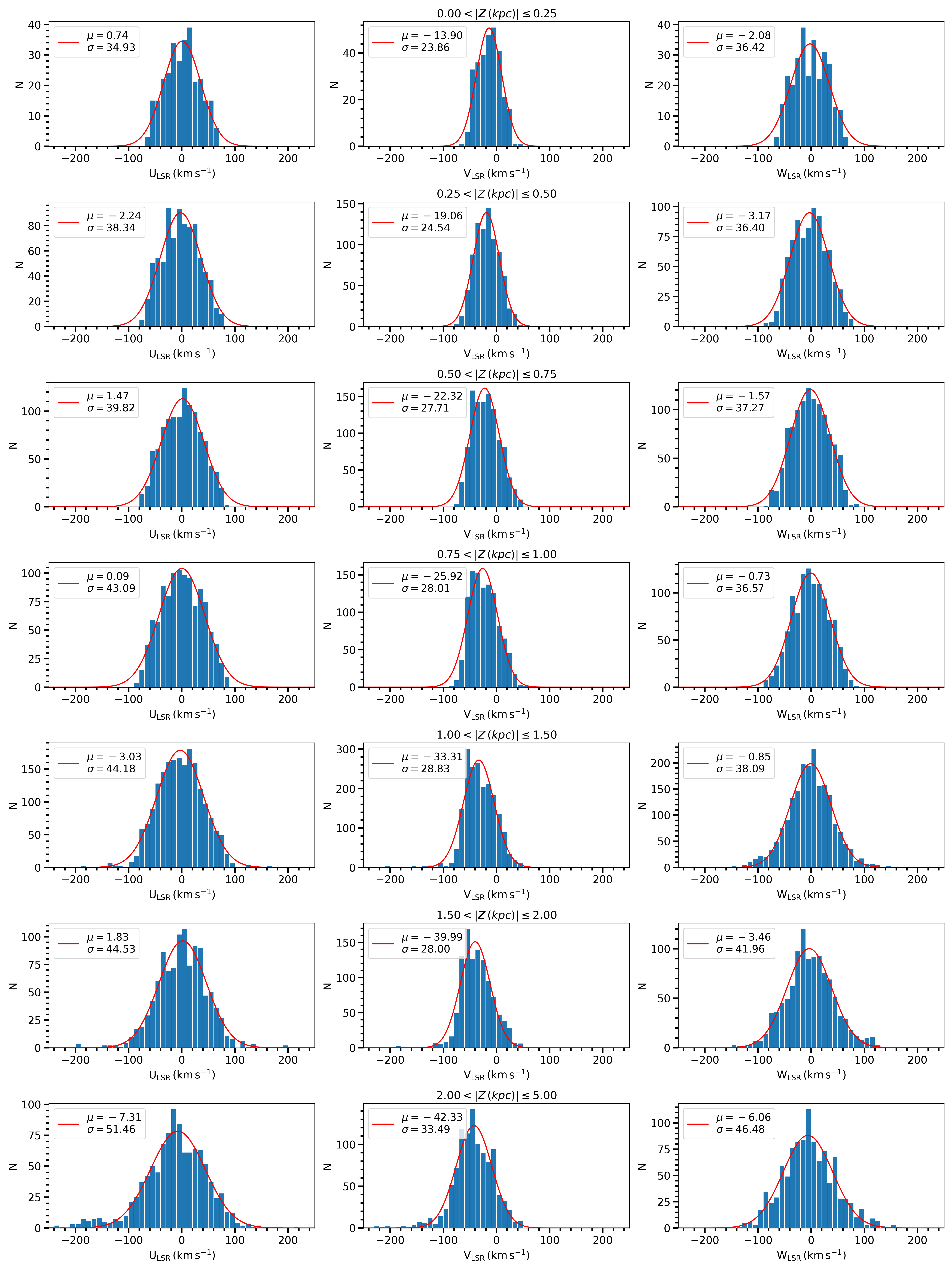}
\caption{Velocity histograms for high-$\alpha$ population for different $Z$ distance intervals. Blue bars are the velocity distributions and red solid lines are the Gaussian fits to distributions in each distance interval.}
\label{fig:Fig14}
\end {figure}

For the high-$\alpha$ sample, all of the space velocity components vary in a larger range with respect to the low-$\alpha$ analogues. $V_{\rm LSR}$ values fit to the thin disc kinematics up to 1 kpc, then they start to obey the thick disc kinematics \citep[i.e.][]{Chiba00} and $V_{\rm LSR}$ reaches to -42.33 km s$^{-1}$ at $2<|Z|\rm \rm (kpc)<5$ \citep[i.e.][]{Soubiran03, Allende16}. Distribution of each velocity component for APOGEE-2 RGB stars in each distance interval is given in Fig. \ref{fig:Fig14}. Based on the figure, velocity distribution in $U_{\rm LSR}$ fits to the Gaussian distribution up to 1 kpc, where classical thick disc kinematics become dominant, i.e. $0.75<|Z|(\rm kpc)<1$ and $1<|Z|(\rm kpc)<1.5$ intervals. Then, the wings of the distribution start to spread after the $1<|Z|(\rm kpc)<1.5$ interval. A midly apparent population of stars at $-200<U_{\rm LSR} (\rm km ~s^{-1})<-100$ become prominent in the $2<|Z|\rm (kpc)<5$ interval. $V_{\rm LSR}$ distribution is generally right-skewed and at $1<|Z|(\rm kpc)<1.5$ interval the left wing of the distribution starts to fill by a population with $-200<V_{\rm LSR} (\rm km~s^{-1})<-100$. Dispersions of each space velocity component for the high-$\alpha$ population, $(\sigma_ {\rm U},\sigma_{\rm V},\sigma_{\rm W})=(42,28,39)$ km s$^{-1}$, are in good agreement with the values for the high-$\alpha$ component in the Solar cylinder of \citet{Allende16}, $(\sigma_R,\sigma_{\phi},\sigma_Z)=(40,34,39)$ km s$^{-1}$. Difference between $\sigma_R$ and $\sigma_U$ or $\sigma_V$ and $\sigma_{\phi}$ values might be caused either by the difference of the definition of the coordinates in heliocentric and Galactic coordinate system or the differential rotation corrections. Metallicity and space velocity component  distributions of RGB stars in high-$\alpha$ population are shown for each distance interval in Fig. \ref{fig:Fig13} and Fig. \ref{fig:Fig14}. Distributions of [Fe/H] in $0<|Z|\rm (kpc)<0.25$ and $0.25<|Z|\rm (kpc)<0.5$ intervals are slightly not fit to the Gaussian curve. Also, it can be seen from the figure there are contributions to the metal-rich and metal-poor wings in distributions. Metallicity distribution follows the normal distribution at $0.5<|Z|\rm (kpc)<0.75$ interval. This normal distribution is accompanied with the small contributions to the metal-poor tail of the distribution in the remaining distance intervals up to $|Z|>2$ kpc. 

\begin{figure}
\centering
\includegraphics[width=\columnwidth]{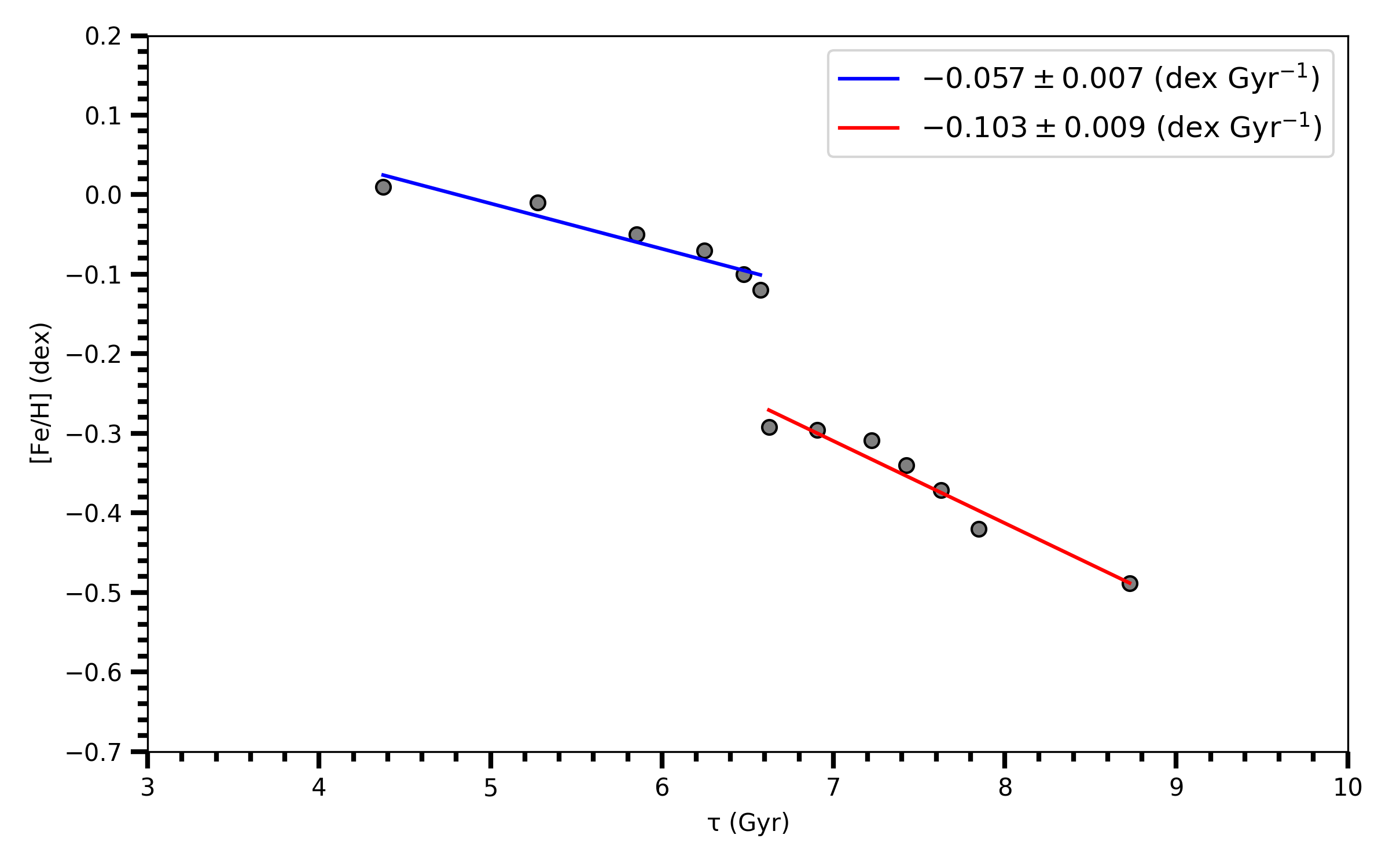}
\caption{Age-metallicity relation for thin (blue line) and thick (red line) disc populations.}
\label{fig:Fig15}
\end {figure} 

Although Galactic disc components have distinct properties in kinematics, chemistry and age, they appear as spatially intertwined structures on $Z\times R_{\rm gc}$ plane, which can be seen from the Fig. \ref{fig:Fig9}. This means that there are some stars with the thick disc chemistry and the thin disc kinematics and some other stars with vice versa. Some of these are old thin disc stars and some are not. Table 1 indicates that regardless of their superposed appearance on the spatial plane, RGB stars have distinct chemistry and age distribution. Also, based on the total space velocity dispersions both components also have distinct characteristics.

Overall results show both Galactic disc components have AMR relations (see Fig. \ref{fig:Fig15}). AMR are calculated via locii points that represent the $|Z|$ distance intervals for each Galactic disc component. Each locus point is determined from the age and metallicity parameters of stars in the respective distance interval. Then the linear fits are applied on all of locii points for the thin and thick discs, separately. Blue and red solid lines represents the low- and high-$\alpha$ populations, respectively. The calculated AMR slopes are-0.057$\pm$0.007 dex Gyr$^{-1}$ for the low-$\alpha$ population, while it is -0.103$\pm$0.009 dex Gyr$^{-1}$ for the high-$\alpha$ population. This study deviates from the literature of AMR derivations a group age from the kinematic data instead of assigning an isochrone-based statistical age from derived observational parameters \citep[e.g.][]{Haywood13, Feuillet19}. Nonetheless our results can be compared with some studies. In one study \citet{Haywood13} have determined the AMR for thin and thick discs using the HARPS spectroscopy, re-reduced \textit{Hipparcos} parallaxes \citep{vanLeeuwen07} and isochrone based ages of 1,111 FGK stars. Their result for the thin disc is -0.025 dex Gyr$^{-1}$ and for the thick disc is -0.15 dex Gyr$^{-1}$. \cite{Haywood13} calculated the enrichment rate for the thick disc as around 5-6 times the thin disc. Even though our results seem to be similar in trend to \citet{Haywood13}, the calculated enrichment rates reflects a difference. We found the high-$\alpha$ enrichment rate as $\sim$1.8 times that of the low-$\alpha$. Vertical velocity dispersions of low- and high-$\alpha$ populations correlate with kinematic age. In Fig. \ref{fig:Fig15}, the first locus of the low-$\alpha$ sample is around the same spot as the Sun, which is [Fe/H]=0 dex and $\tau=4.6$ Gyr. So, we cannot provide kinematic age of the stellar group more rich in metallicity based on this analysis even though our sample reaches to +0.5 dex at the metal-rich tail of the metallicity distribution. 

The oldest star group in the low-$\alpha$ population has a mean age of 6.57 Gyr and a mean metallicity of -0.12 dex with rotational velocities -13.42 km s$^{-1}$. On the other hand for the high-$\alpha$ population, the youngest group have a mean age of 6.62 Gyr, metallicity of -0.29 dex and the rotational velocity of -13.90 km s$^{-1}$. Both samples are around the same age and similar in kinematics, but the metallicity differs by 0.15 dex. Difference in chemistry implies that these samples presumably have different origin, and their rotational velocities are almost same. It's because kinematic properties of stars have tendency to adapt the orbital angular momentum distribution that is shaped by the large scale perturbation sources.

So, there is a kinematic continuity within the Galactic discs, of the regardless the chemical sub-populations and a discontinuity among the chemical populations. However, each chemical group has their own chemical continuities from metal rich to metal poor as the vertical distances increase from the Galactic plane. It can be inferred from this result to chemical discs have different origins related to the gas accretion.

\section{Summary and Conclusion}
We have investigated the age-metallicity relation for the Galactic disc using the kinematic properties calculated from the precise astrometric and spectroscopic data of 43,592 RGB stars within the Solar cylinder. Red giant sample is selected using PARSEC stellar evolution tracks. Relative parallax errors are limited to 0.1 and Lutz-Kelker bias is not applied on the trigonometric parallaxes of the sample. Stellar distances are determined via inverse parallax relation. Based on the difference between inverse parallax method and BJ18 methods compared and we shown that the estimation method is not a crucial tool up to 2 kpc distances. Precise input parameter in kinematic calculations allowed more dependable ages than the ones obtained with the former kinematics data. In the study, Galactic populations are defined using a GMM  applied to the $[\alpha/{\rm Fe}]\times$[Fe/H] plane, see Sect. 3.1. As a result of this method, chemical disk sub-components emerged as low-$\alpha$ and high-$\alpha$ populations, which are akin to the classical thin disc and thick disc/halo, respectively. Each chemical disc component was divided into $|Z|$ intervals, and they became the locii points for AMR calculation. 

This study shows there are AMRs for both disc components. Past studies usually found an AMR for the thick disc and no relation for the thin disc due to spread in metallicity. As both the quality and the quantity of the photometric, spectroscopic and the astrometric data are enhanced within the last decade, relations between stellar age-metallicity and stellar age-various elemental ratios immensely investigated with similar, but different methodologies. Moreover, individual stellar age plays the key role on the modelling efforts. We hope the PLAnetary Transits and Oscillation of Stars mission (PLATO) \citep{Miglio17} shall remove most of the fuzzy parts of the Galactic formation and evolution models by providing excellent age determinations. From the Galactic archaeology point of view, PLATO survey is expected to derive astroseismic ages for red giant stars with a 10\% precision.


\section{Acknowledgements}
We thank the anonymous referee for his/her insightful and constructive suggestions, which significantly improved the paper. This study is related to Sibel D\"oner's MSc thesis. This work has been supported by the Scientific and Technological Research Council of Turkey (T\"UB\. ITAK), Grant No: MFAG-118F350. 
We have also made use of data from the APOGEE survey, which is part of Sloan Digital Sky Survey IV. SDSS-V is managed by the Astrophysical Research Consortium for the Participating Institutions of the SDSS Collaboration (http: //www.sdss.org). This work has made use of data from the European Space Agency (ESA) mission {\it Gaia} (\url{https://www.cosmos.esa.int/gaia}), processed by the {\it Gaia}Data Processing and Analysis Consortium (DPAC, \url{https://www.cosmos.esa.int/web/gaia/dpac/consortium}). Funding for the DPAC has been provided by national institutions, in particular the institutions participating in the {\it Gaia} Multilateral Agreement. This research has made use of NASA’s Astrophysics Data System Bibliographic Services.



\bibliographystyle{mnras}
\bibliography{biblio_list}

\bsp	
\label{lastpage}
\end{document}